\documentclass[
superscriptaddress,
twocolumn,
amsmath,amssymb,
aps,
prx,
nofootinbib,
]{revtex4-2}

\usepackage{comment}
\usepackage{tabularx}
\usepackage{graphicx}
\usepackage{dcolumn}
\usepackage{bm}
\usepackage{mathtools}
\usepackage{color}

\usepackage{hyperref}
\hypersetup{
	colorlinks=true,
	linkcolor=blue,
	citecolor=blue,
	urlcolor=blue
}

\usepackage{dsfont}

\newcommand{\E}{\mathcal{E}}

\begin{document}

\title{Quantum-State-Controlled Collisions of Ultracold Polyatomic Molecules}

\author{Nathaniel B. Vilas}
\email{vilas@g.harvard.edu}
\author{Paige Robichaud}
\author{Christian Hallas}
\author{Junheng Tao}
\affiliation{Department of Physics, Harvard University, Cambridge, MA 02138, USA}
\affiliation{Harvard-MIT Center for Ultracold Atoms, Cambridge, MA 02138, USA}
\author{Lo\"ic Anderegg}
\altaffiliation{Present address: Department of Physics and Astronomy, University of Southern California, Los Angeles, CA 90089, USA}
\affiliation{Department of Physics, Harvard University, Cambridge, MA 02138, USA}
\affiliation{Harvard-MIT Center for Ultracold Atoms, Cambridge, MA 02138, USA}
\author{Grace K. Li}
\author{Hana Lampson}
\affiliation{Department of Physics, Harvard University, Cambridge, MA 02138, USA}
\affiliation{Harvard-MIT Center for Ultracold Atoms, Cambridge, MA 02138, USA}
\author{Lucie D. Augustovi\v{c}ov\'{a}}
\affiliation{Charles University, Faculty of Mathematics and Physics, Department of Chemical Physics and Optics, Ke Karlovu 3, CZ-12116 Prague 2, Czech Republic}
\author{John L. Bohn}
\affiliation{JILA, NIST, and Department of Physics, University of Colorado, Boulder, Colorado 80309, USA}
\author{John M. Doyle}
\affiliation{Department of Physics, Harvard University, Cambridge, MA 02138, USA}
\affiliation{Harvard-MIT Center for Ultracold Atoms, Cambridge, MA 02138, USA}

\date{\today}

\begin{abstract}

Collisions between ultracold calcium monohydroxide (CaOH) molecules are realized and studied. Inelastic collision rate constants are measured for CaOH prepared in ground and excited vibrational states, and the electric field dependence of these rates is measured for molecules in single quantum states of the parity-doubled bending mode. Theoretical calculations of collision rate coefficients are performed and found to agree with measured values. The lowest collisional loss rates are for states with repulsive long-range potentials that shield ultracold molecules from loss channels at short distance.
These results unveil the collisional behavior of parity doublet molecules in the ultracold regime, and lay the foundation for future experiments to evaporatively cool polyatomic molecules to quantum degeneracy.

\end{abstract}

\maketitle

\section{Introduction}

Ultracold molecules are promising for a diverse range of applications,
including searches for physics beyond the Standard Model~\cite{safronova2018search, kozyryev2017precision, kozyryev2021enhanced}, quantum simulation and information science~\cite{demille2002quantum, micheli2006toolbox, yu2019scalable}, and studies of cold and ultracold chemistry~\cite{heazlewood2021towards}. Progress over the last decade with diatomic molecules has led to rapid advancements in direct cooling~\cite{barry2014magneto, anderegg2017radio, truppe2017molecules, collopy20183d, zeng2024three}, molecule assembly~\cite{ni2008high, takekoshi2014ultracold, park2015ultracold, guo2016creation, rvachov2017long, cairncross2021assembly}, optical trapping~\cite{anderegg2018laser, anderegg2019optical, zhang2022optical}, single-state control~\cite{chou2017preparation, park2017second, Burchesky2021}, quantum entanglement~\cite{yan2013observation,christakis2023probing, holland2023on, bao2023dipolar, miller2024two, carroll2025observation, picard2025entanglement, ruttley2025long}, and studies of ultracold collisions~\cite{ospelkaus2010quantum,ni2010dipolar,guo2018dipolar,ye2018collisions,gregory2019sticky,cheuk2020observation,yan2020resonant,park2023feshbach,burau2024collisions}.
Experiments have measured collisional inelastic loss rates that are typically near the universal limit,
in which the probability of loss after a collision at short range is unity~\cite{idziaszek2010universal,ospelkaus2010quantum,ye2018collisions,gregory2019sticky,cheuk2020observation,burau2024collisions, bause2023ultracold}.
However, it has also been demonstrated that electric dipole interactions and applied fields can be used to engineer repulsive potentials that shield the molecules from short-range loss~\cite{valtolina2020dipolar,matsuda2020resonant,li2021tuning,anderegg2021observation}.
These techniques have enabled
diatomic molecules to be evaporatively cooled
to quantum degeneracy~\cite{valtolina2020dipolar,Schindewolf2022,bigagli2024observation}.

Polyatomic molecules are a newer frontier,
with small polyatomic species having recently been
laser cooled to microkelvin temperatures~\cite{vilas2022magneto}, loaded into optical traps~\cite{hallas2023optical} and tweezer arrays~\cite{vilas2024optical}, and controlled at the single quantum state level~\cite{anderegg2023quantum,vilas2024optical, low2025coherence}.
The diverse structures present in polyatomic molecules provide new opportunities for research in
quantum information science~\cite{wei2011entanglement,yu2019scalable, albert2020robust}, quantum simulation~\cite{wall2013simulating,wall2015realizing}, ultracold chemistry~\cite{heazlewood2021towards}, and precision searches for physics beyond the Standard Model~\cite{kozyryev2017precision,kozyryev2021enhanced}.
Polyatomic molecules generically contain structures, most notably the presence of closely-spaced states of opposite parity~\cite{hutzler2020polyatomic,augenbraun2023review}, that are expected to result in qualitatively distinct collisional properties~\cite{avdeenkov2002collisional,avdeenkov2003linking,avdeenkov2005ultracold, augustovicova2019collisions}.
Examples include electrostatic shielding methods for evaporative cooling~\cite{augustovicova2018no,augustovicova2019collisions}, and field-linked states that can be used to tune collision rates and assemble larger polyatomic molecules~\cite{avdeenkov2003linking, augustovicova2019collisions,chen2023field,chen2024ultracold}.
Experimentally, collisions between polyatomic molecules have been previously observed in the $0.1-1$~K temperature regime~\cite{wu2017cryofuge,koller2022electric}, where dipolar relaxation was the dominant collisional effect~\cite{koller2022electric}.
Despite decades of theoretical studies~\cite{bohn2001inelastic,avdeenkov2002collisional,avdeenkov2003linking,avdeenkov2005ultracold,augustovicova2018no,augustovicova2019collisions}, collisions between molecules with parity doublet structure, including polyatomic species, have not previously been observed at ultracold temperatures ($<$1~mK).

Here, we  report on the observation of collisions between optically trapped, bosonic CaOH molecules at temperatures $\sim$100~$\mu$K.
Compared to previous experiments at $\gtrsim$0.1~K, the molecules are colder than all rotational energy scales, parity doublet splittings included, 
and are in the few-partial-wave regime.
We observe inelastic collisional loss for
CaOH prepared in both the vibrational ground state and in the lowest-energy vibrational bending mode. The measured collisional loss rate in the bending mode is higher than in the ground state, which we attribute to the parity doublet structure of the bending mode.
We also measure collisional loss rates of molecules in single quantum states in the bending mode, including as a function of applied electric field.
We compare the measurements to calculations that include long-range dipolar interactions and universal short-range loss, and use these comparisons to build physical understanding of the collisional behavior of parity-doublet molecules.
Certain parity-doublet states are identified that have repulsive long-range interaction potentials, which are calculated to have high ratios of elastic to inelastic collision rates at lower temperature.

\section{Experimental protocol}

The starting point of the experiment is a conveyor-belt blue-detuned magneto-optical trap (MOT) of CaOH~\cite{hallas2024high, li2025conveyor}. Compared to previously realized red-detuned MOTs of CaOH~\cite{vilas2022magneto, hallas2023optical}, the conveyor MOT has approximately two orders of magnitude higher density, directly enabling ultracold collision experiments for laser-cooled polyatomic molecules. After forming the conveyor MOT, CaOH molecules are then loaded
into an optical dipole trap (ODT), which is formed from a 1064~nm laser with a Gaussian beam waist of $\sim$9~$\mu$m and a trap depth of $\sim$650~$\mu$K.
The number of molecules initially loaded into the ODT is measured using a short ($\leq5$~ms), nondestructive \emph{in situ} single-frequency (SF) imaging pulse~\cite{hallas2023optical}, followed by a 2~ms recooling pulse. Next, the molecules are prepared in the desired internal state (using a process described later in the text).

To increase the density so that collisions occur on a faster timescale than single-molecule loss (which is due to thermalization of vibrational state population~\cite{hallas2023optical,vilas2023blackbody}), the trap is adiabatically ramped up to a trap depth ($U_0$) of $U_0/k_B \approx 4.5$~mK.
The initial temperature of the molecules after the ramp is $T_0 = 80$-$100~\mu$K, which depends on the prepared internal state. The trap frequencies are $\{\omega_x, \omega_y, \omega_z\} = 2\pi \times \{30.4,26.2,0.55\}$~kHz, corresponding to a peak number density of
$n_0 \gtrsim 10^{11}$~cm$^{-3}$ for the approximately 100 molecules in the trap~\cite{Supplemental}.

At high density in the trap, molecules undergo elastic collisions, inelastic state-changing collisions, and short-range losses (e.g. chemical reactions~\cite{ospelkaus2010quantum,augustovicova2019collisions} or complex formation~\cite{bause2023ultracold}).
Elastic collisions thermalize the molecules.
At the ratio of the trap depth to the molecule temperature in the experiment, $\eta = U_0/k_B T_0 \approx 45$, all observed collisional losses will be from inelastic and short-range processes (for brevity, we will refer to both of these processes as ``inelastic'').
Our experiments are designed to only detect molecules remaining in the initial state, meaning that all inelastic collisions will appear as loss in our data (some inelastic collisions will also impart enough kinetic energy to physically eject molecules from the trap).

After holding the molecules for a variable amount of time, the surviving molecules are detected by switching off the trap and imaging the cloud in free space using a 40~ms pulse of $\Lambda$-cooling light~\cite{cheuk2018lambda, hallas2023optical}.
The imaging is performed in free space to avoid systematic effects due to light-assisted collisions in the trap.
The experimental sequence is described in more detail in the Supplemental Material~\cite{Supplemental}.

\section{Observation of molecule-molecule collisions}

\begin{figure}
    \centering
    \includegraphics{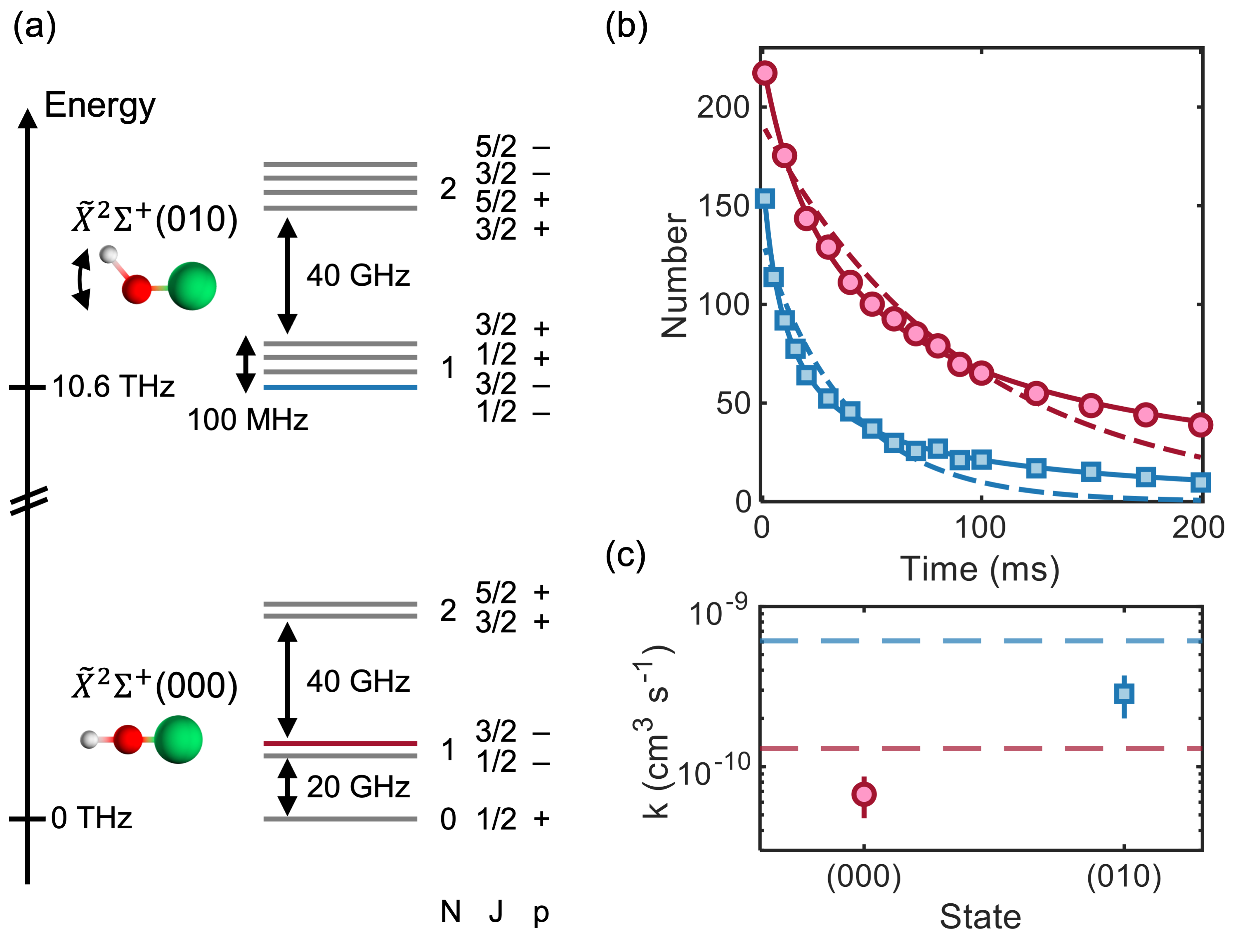}
    \caption{Observation of collisions between optically trapped CaOH molecules. (a) Rotational structure of CaOH in the vibrational ground state, $\widetilde{X}(000)$, and vibrational bending mode, $\widetilde{X}(010)$. Opposite-parity states, which must be mixed to turn on dipolar interactions, are spaced by approx. 20~GHz in the ground state and by approx. 40~MHz in the bending mode. States are labeled by the rotational quantum number $N$, total angular momentum (excluding nuclear spin) $J$, and parity $p$. (b) Molecule number vs. hold time in the ODT for molecules prepared at $T=80$~$\mu$K in $\widetilde{X}(000)(N=1, J=3/2)$ (red circles) and in $\widetilde{X}(010)(N=1, J=1/2^-)$ (blue squares). Solid curves are fits to the two-body collision model described in the text, and dashed curves are fits to a single exponential timescale. (c) Fitted collisional loss rate constants, $k$, for the two vibrational states. Dashed lines are the universal loss rates calculated for these states~\cite{Supplemental}.
    }
    \label{fig:1}
\end{figure}

Fig.~\ref{fig:1}(a) depicts the low-lying rotational and vibrational states of CaOH in the $\widetilde{X}^2\Sigma^+$ ground electronic potential. In our first set of experiments, we prepare CaOH molecules in a mixture of hyperfine states within a single rotational state. This is done, separately, for molecules in either the ground vibrational state [$\widetilde{X}(000)$] or the vibrational bending mode [$\widetilde{X}(010)$]. Vibrational states are labeled with $(v_1 v_2 v_3)$, where $v_1$ is the vibrational quantum number for the Ca--O stretching mode, $v_2$ is for the bending mode, and $v_3$ is for the O--H stretching mode. 
Each rotational level in the $\widetilde{X}(010)$ bending mode contains opposite parity (parity-doublet) states spaced by tens of MHz, which
correspond to bending vibrations in two perpendicular planes in the molecule frame. Their degeneracy is broken by Coriolis interactions~\cite{hutzler2020polyatomic}.

Fig.~\ref{fig:1}(b) shows the molecule number as a function of hold time for molecules prepared in either the $\widetilde{X}(000)(N=1, J=3/2^-)$ level of the vibrational ground state (red circles) or the $\widetilde{X}(010)(N=1, J=1/2^-)$ level of the vibrational bending mode (blue squares)~\cite{Supplemental}. In both cases, the molecules are in an incoherent mixture of hyperfine states arising from the hydrogen nuclear spin $I=1/2$. The data fit very well to two-body decay curves (solid lines) while fitting very poorly to exponential one-body curves (dashed lines), showing that CaOH-CaOH collisions dominate.

To determine the CaOH-CaOH collision rate constant, we fit the data to a rate equation model that includes the two-body loss rate constant $k$ and the single-molecule background lifetime $\tau$. We also account for molecular heating due to a mechanism where the coldest molecules are the most likely to collide and be lost from the trap. The coupled rate equations for the molecule number $N$ and temperature $T$ are~\cite{gregory2019sticky}
\begin{subequations}
\begin{align}
\frac{dN(t)}{dt} &= -k \frac{1}{V_\text{eff}(T)} N(t)^2 - \frac{1}{\tau}N(t) \\
\frac{dT(t)}{dt} &= \frac{1}{4} k \frac{1}{V_\text{eff}(T)}N(t)T(t)
\end{align}
\label{eqn:rateequations}
\end{subequations}
where $V_\text{eff}(T) = \xi(T)(4\pi k_B T/m)^{3/2}(\omega_x \omega_y \omega_z)^{-1}$ is the effective trap volume, and $\xi(T)\approx 1.1$ is a factor that accounts for the finite temperature of the molecules~\cite{Supplemental}.
The single-molecule lifetimes $\tau$ are much longer than the observed collision timescales, and are held
fixed in the fit to the measured values of $\tau_{000} \approx 900$~ms and $\tau_{010} \approx 360$~ms for the ground state and bending mode, respectively~\cite{hallas2023optical,vilas2023blackbody}.
Fits to the rate model of eqn.~\ref{eqn:rateequations} are shown as solid curves in Fig.~\ref{fig:1}(b). The measured collision rate constants are $k_{(000)}=0.7(2) \times 10^{-10}$~cm$^{3}$~s$^{-1}$ and $k_{(010)}=2.9(9) \times 10^{-10}$~cm$^{3}$~s$^{-1}$, plotted in Fig.~\ref{fig:1}(c).
These values include a small ($\sim$10\% level) correction that accounts for collisions with
molecules trapped in the wrong state~\cite{Supplemental}.

The inelastic loss rate constant $k$ for molecules in the bending mode is $\sim4\times$ higher than in the vibrational ground state. We attribute this difference to an increased
interaction strength from the parity doublet structure in the bending mode.
Molecules interact via a van der Waals (vdW) potential $V(r) = -C_6/r^6$ arising from electric dipole interactions, wherein the electric field from one molecule induces a dipole moment in the second molecule by mixing states of opposite parity. In the ground state, the nearest opposite-parity states are rotational levels split by the rotational constant $B = h\times 10.023$~GHz, while in the bending mode the opposite-parity states are split by the $\ell$-doubling parameter $q_\ell \approx h\times 21.5$~MHz. Ignoring electronic contributions, in second order perturbation theory the vdW coefficient is $C_{6,(000)} \approx \frac{1}{6B}\frac{d^4}{(4\pi \epsilon_0)^2}$ in the ground state and $C_{6,(010)} \approx 0.85 \times \frac{1}{24q_\ell}\frac{d^4}{(4\pi \epsilon_0)^2}$ in $J=1/2^-$ of the bending mode, where $d = 1.465$~D is the molecule frame dipole moment~\cite{steimle1992supersonic} and the numerical prefactor accounts for substructure from the electron spin~\cite{Supplemental}. The closely-spaced parity doublets therefore increase the van der Waals interaction strength by $C_{6,(010)}/C_{6,(000)} \sim B/q_\ell \sim 10^2$.

In the universal limit, where molecules are lost whenever they collide at short range (for instance due to the chemical reaction CaOH + CaOH $\rightarrow$ Ca(OH)$_2$ + Ca~\cite{augustovicova2019collisions}), $k$ is expected to be proportional to the van der Waals length $r_\text{vdW} = \frac{1}{2}(2 \mu C_6/\hbar^2)^{1/4}$ at low temperature~\cite{idziaszek2010universal}, where $\mu= (57~\text{u})/2$ is the reduced mass. Using the $C_6$ values above, we estimate $r_{\text{vdW},(000)} \approx 94~a_0$ and $r_{\text{vdW},(010)} \approx 297~a_0$,
implying a $\sim$$3\times$ higher collisional loss rate in the bending mode compared to the ground state.
However, at the temperatures in our experiment,
the universal loss rate deviates slightly from this prediction and is consistent with a classical Langevin capture model~\cite{bell2009ultracold,jurgilas_2021}.
The calculated universal loss rates
at $T=80$~$\mu$K are $k^\text{univ}_{(000)}=1.3 \times 10^{-10}$~cm$^3$~s$^{-1}$ and $k^\text{univ}_{(010)}=6.1 \times 10^{-10}$~cm$^3$~s$^{-1}$, shown as horizontal lines in Fig.~\ref{fig:1}(c)~\cite{Supplemental}. The calculated universal rate is $\sim$$4.5\times$ higher in the bending mode than in the ground state, and both measured rates are approximately 50\% below the universal limit. Details on the universal loss rate calculations are provided in the Supplemental Material~\cite{Supplemental}.

\section{Dependence on internal state and electric field}

\begin{figure*}
    \centering
    \includegraphics{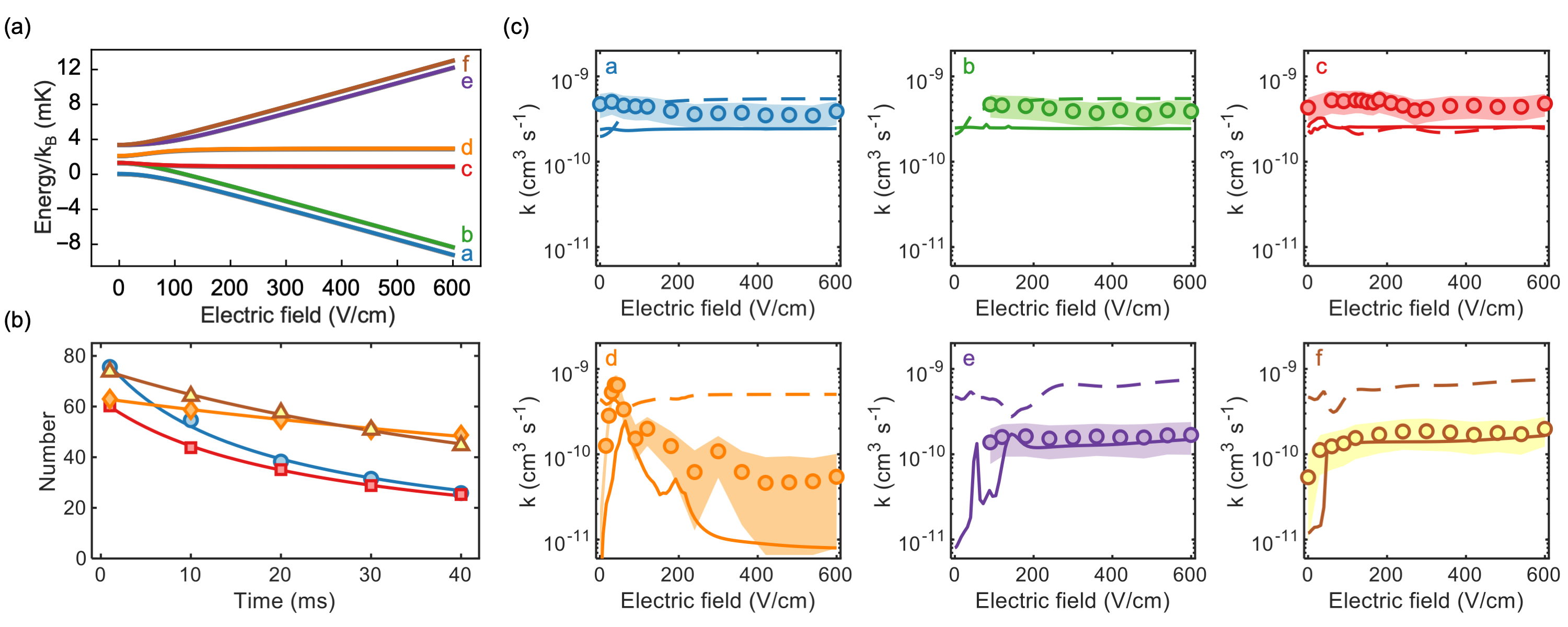}
    \caption{Electric field dependence of CaOH-CaOH collisions in the bending mode. (a) Single-molecule energies in the $\widetilde{X}(010), N=1$ manifold of CaOH. At electric fields $\E\gtrsim 150$~V/cm, opposite parity-doublet states are mixed and the molecular dipole moment is aligned in the laboratory frame. In this regime, the structure separates into six manifolds, labeled $a$-$f$. (b) Collision lifetime curves for molecules prepared at $\E=0$ in a single hyperfine state in the $a$ (blue circles), $c$ (red squares), $d$ (orange diamonds), and $f$ (brown triangles) manifold. Solid curves are fits to the rate equation model used to extract the two-body loss rate constant $k$. (c) Measured collisional loss rate constants $k$ vs. electric field for single hyperfine states in the $a$-$f$ manifolds. The shaded regions denote the experimental uncertainty (standard error), which accounts for statistical error, systematic uncertainty in the molecule number density, and uncertainty in the background loss rate from imperfect state preparation~\cite{Supplemental}. Solid curves are calculated loss rate coefficients assuming universal short range loss, and dashed curves are the calculated elastic collision rate coefficients.}
    \label{fig:2}
\end{figure*}

To further investigate the effect of parity doublet structures on CaOH-CaOH collisions,
we next prepare the molecules in single quantum states within the $N=1$ rotational manifold in the bending mode, and then tune the interaction potential by applying an external electric field. The structure of this manifold is shown as a function of applied electric field in Fig.~\ref{fig:2}(a). At zero field, there are four well-resolved levels ($J=1/2^-$, $J=3/2^-$, $J=1/2^+$, and $J=3/2^+$ in order of increasing energy). These are split by the spin-rotation and $\ell$-type parity doubling interactions. Under the presence of an applied electric field $\E$, the opposite-parity states are mixed. For $\E\gtrsim150$~V/cm, the structure splits into 6 well resolved manifolds, which we label $a$-$f$ in order of increasing energy~\cite{augustovicova2019collisions}. In this ``high-field'' regime, the molecular dipole moment is aligned in the laboratory frame.
Note that each manifold contains several hyperfine states, split by energies of $\lesssim k_B \times 100$~$\mu$K, due to the hydrogen nuclear spin.

To prepare single quantum states, we first optically pump the molecules into the $(J=1/2^-, F=0)$ hyperfine state~\cite{anderegg2023quantum, vilas2024optical},
then use microwave and radio-frequency pulses to coherently transfer population between manifolds. For each manifold, population is prepared in the highest-energy hyperfine state and with a small applied magnetic field ($B=0-3$~G depending on the state), which ensures that the state does not undergo level crossings when adiabatically ramping the electric field.
We measure collision rates for electric fields in the range $\E=0-600$~V/cm by
recording the number of molecules vs. hold time at high density and fitting to eqn.~\ref{eqn:rateequations} (Fig.~\ref{fig:2}(b)).
To minimize redistribution of population into impurity states driven by blackbody radiation (BBR)~\cite{vilas2023blackbody}, we record data only for hold times $\lesssim$40~ms, which is significantly less than the BBR-limited bending mode lifetime. We detect only the molecules in the target collision state, and push most of the molecules in other states out of the trap at several points during the sequence.
Approximately 80\% of the molecules in the trap during the collision time are prepared in the correct hyperfine state, with the remaining molecules being in undetectable rovibrational states due to BBR- and laser cooling-induced losses.
See the Supplemental Material~\cite{Supplemental} for more details on the state preparation and experimental sequence.

Fig.~\ref{fig:2}(c) shows measured collisional loss rate constants for the $a$-$f$ manifolds as a function of applied electric field. Experimental uncertainties are denoted by shaded regions.
At high collision rates, the error is dominated by uncertainty in the trapped molecule density~\cite{Supplemental}.
At low collision rates, the dominant uncertainty is from undesired collisions with molecules trapped in dark rovibrational states.
The effective rate of these ``background'' collisions is estimated to be $k_\text{bg} \approx 7(3) \times 10^{-11}$~cm$^3$~s$^{-1}$~\cite{Supplemental}, and likely depends on the state and electric field. We subtract $k_\text{bg}$ from the fitted rate constants to obtain the rates shown in Fig.~\ref{fig:2}(c), with a corresponding uncertainty accounted for in the shaded region.

We also perform close-coupling calculations of the rate coefficients accounting for long-range dipolar interactions and universal short-range loss, at a collision energy $E_c = k_B \times 100$~$\mu$K. Details are provided in the Supplemental Material~\cite{Supplemental}. Calculated loss rate coefficients $k$ are plotted as solid curves in Fig.~\ref{fig:2}(c), and the corresponding elastic collision rate coefficients $k_\text{el}$ are shown as dashed curves. We observe generally good agreement between the measurements and the calculations, within experimental uncertainty.

\section{Discussion}

\begin{figure}
    \centering
    \includegraphics{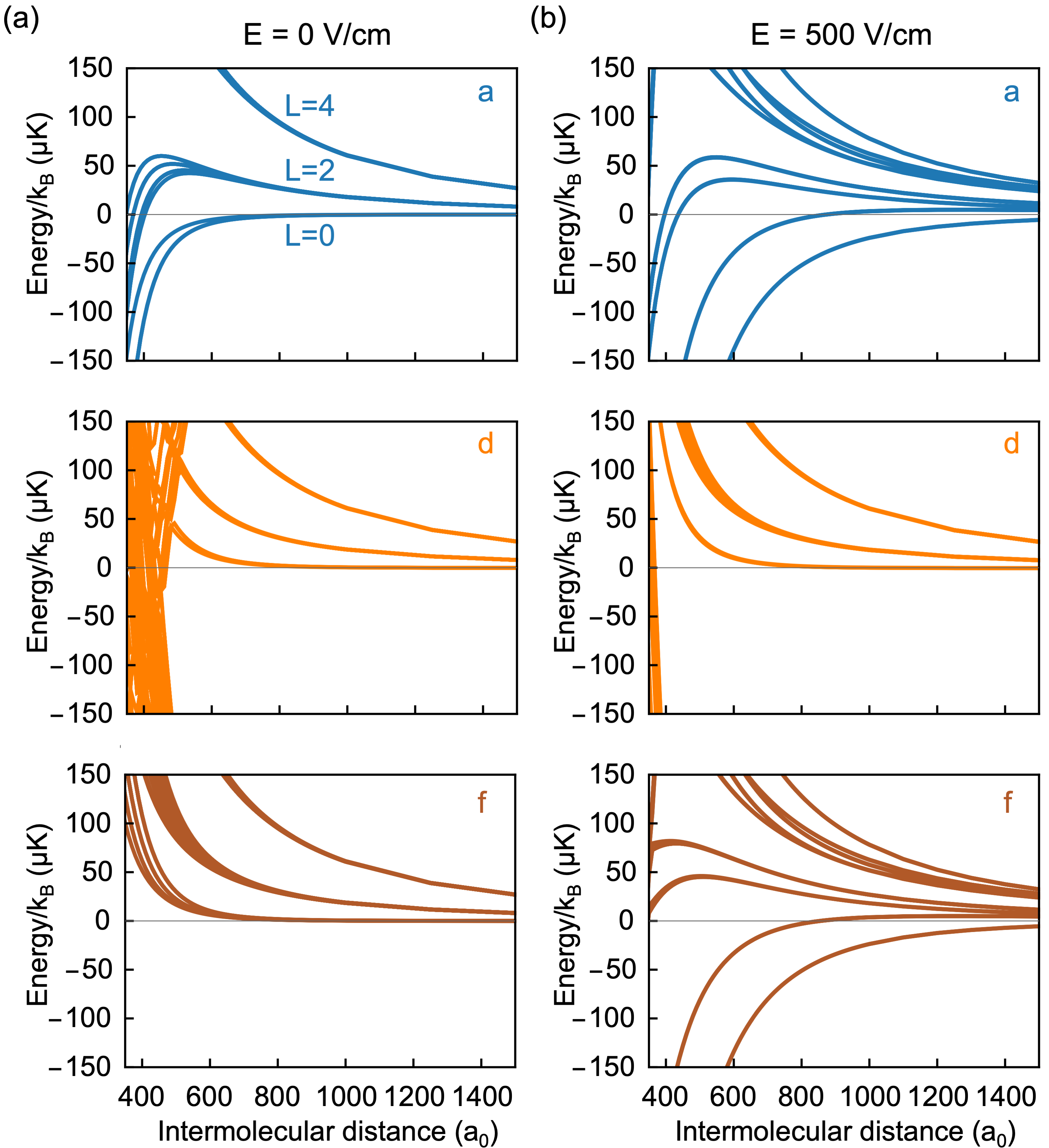}
    \caption{Adiabatic dipolar interaction potentials for the $a$, $d$, and $f$ states at (a) $\mathcal{E}=0$~V/cm and (d) $\mathcal{E}=500$~V/cm. The first three even partial waves ($L=0,2,4$) are shown. To reduce the density of states and improve visual clarity, these calculations do not include hyperfine structure, applied magnetic fields, or AC Stark shifts, none of which significantly influence the long-range interactions or qualitatively alter the shape of the potentials.}
    \label{fig:3}
\end{figure}

Several general features are apparent
in the data of Fig.~\ref{fig:2}(c).
The collision rates vary primarily over the electric field range $\E=0-150$~V/cm, then plateau at higher fields once the molecules are aligned in the laboratory frame.
The $a$-$c$ manifolds have higher collisional loss rates than the $d$-$f$ states at nearly all fields. The $d$ manifold exhibits a peak in the collision rate around $\E\approx 40$~V/cm, with much lower collision rates on either side, at $\E=0$~V/cm and $\E\gtrsim 300$~V/cm. The rate constants vary by over an order of magnitude across the range of states and electric fields studied.

The observed collisional losses arise from two distinct effects: inelastic relaxation driven by dipolar interactions at long-range~\cite{bohn2001inelastic}, and short-range losses (e.g. chemical reactions), which occur with high probability if the molecules are allowed to collide at short distance ($\lesssim 30~a_0$).
To help develop an intuition for
these effects,
in Fig.~\ref{fig:3} we plot calculated adiabatic molecule-molecule interaction potentials for the $a$, $d$, and $f$ manifolds at $\E=0$~V/cm and $\E=500$~V/cm~\cite{Supplemental}.

At $\E=0$ (Fig.~\ref{fig:3}(a)),
molecules in the $a$ state (as well as $b$ and $c$) experience an attractive potential $-C_6/r^6$, with $C_6$ arising from second-order dipolar mixing with opposite parity states higher in energy, as described above. In contrast, the $d$, $e$, and $f$ manifolds reside in the higher parity manifold, whereby the same mixing induces a repulsive $C_6/r^6$ interaction. The repulsive wall set up in the latter potentials prevents CaOH molecules at $T\approx100$~$\mu$K from reaching separations less than a few hundred Bohr radii ($a_0$), effectively shutting off short-range losses.
The measured collisional loss rates for the $d$ and $f$ manifolds at zero field are $\gtrsim 5-10\times$ lower than the corresponding rates in the $a$ and $c$ manifolds.
The remaining collisional loss in these states is from inelastic relaxation to the negative parity states, driven by dipolar interactions at long range~\cite{bohn2001inelastic, Supplemental}.

At $\E=500$~V/cm (Fig.~\ref{fig:3}(b)), opposite-parity states are mixed by the electric field, causing the molecular dipole moment to be aligned in the laboratory frame.
In the $a$, $b$, $e$ and $f$ states, the dipole moment is aligned (or anti-aligned) with the applied electric field. These states therefore experience first-order dipolar interactions that qualitatively alter the interaction potentials.
Collisional loss rates in these states are relatively high, with contributions from both short-range and inelastic losses.
By contrast, molecules in the $c$ and $d$ states have zero lab-frame dipole moment even at $\E = 500$~V/cm, meaning that their interactions arise from second-order dipolar mixing. The dominant mixing for molecules in the $d$ state is with lower-energy channels,
so these molecules
experience a repulsive $C_6/r^6$ potential (the opposite is true for molecules in the $c$ state). Molecules in the $d$ state are therefore shielded from short-range loss at high field, with the remaining losses dominated by long-range dipolar relaxation.

In the intermediate-field range,
loss rates for the $e$-$f$ states rise rapidly, on the electric field scale over which the dipoles become polarized. This circumstance can be seen qualitatively in two effects, both given in terms of the dimensionless parameter $\beta \sim 2\langle d\E \rangle/ 2q_\ell$, where $2q_\ell$ is the $\ell$-doublet splitting and $\langle d \E \rangle$ represents the matrix element of the Stark Hamiltonian between opposite parity states of the zero-field molecular Hamiltonian~\cite{avdeenkov2002collisional}.
In the CaOH bending mode, $\beta\approx 1$ occurs at an electric field $\E \approx 60$~V/cm.

The first effect of turning on the electric field involves increasing the long-range, inelastic coupling to other scattering channels.
This induced coupling scales as $\beta/(1+\beta^2)(1/r^3)$~\cite{avdeenkov2002collisional}
and is the dominant collisional loss mechanism in the electric field range $\beta \lesssim 1$.
The second effect involves the direct long-range interaction.  Even when the dipoles are partially or fully polarized, the dipole-dipole interaction vanishes in the 
s-wave channel.  Nevertheless, second-order dipolar coupling to the d-wave channels leads to an effective interaction $-C_4/r^4$, with $C_4 \propto (\beta^2/(1+\beta^2))^2$ \cite{avdeenkov2002collisional}. Thus as the molecules become polarized, the effective attraction in this channel grows, encouraging greater incoming flux to reach short range and react chemically.
This effect becomes significant at higher fields ($\beta \gtrsim 1$), at which point the direct interaction strength $C_4$ approaches a constant value while the inelastic coupling falls off as $1/\beta$ (due to the increased spacing between Stark manifolds).

The calculations and measurements for the $d$-$f$ states indeed show that inelastic relaxation initially dominates over short-range absorption, causing collisional loss rates to rise rapidly at fields $\E \sim 0-60$~V/cm as the dipoles become polarized. At higher fields $\E \gtrsim150$~V/cm, the inelastic rate for the $e$-$f$ states decreases while the short-range loss rate grows, and these two rates eventually become comparable. For the $d$ state at high field, inelastic relaxation remains the dominant effect due to the repulsive barrier discussed earlier.
Plots showing the calculated contributions of inelastic and short-range losses to the total loss rate can be found in the Supplemental Material~\cite{Supplemental}.

At intermediate fields, the $d$ state experiences high collisional loss rates due primarily to long-range inelastic relaxation,
which may be enhanced by
level crossings between collision channels
in this range of fields~\cite{Supplemental}.
The result of these many curve crossings appears to be a rich resonant structure revealing quasi-bound states at intermediate separations $r$. Analysis of these resonances remains a task for future investigations.

\begin{figure}
    \centering
    \includegraphics{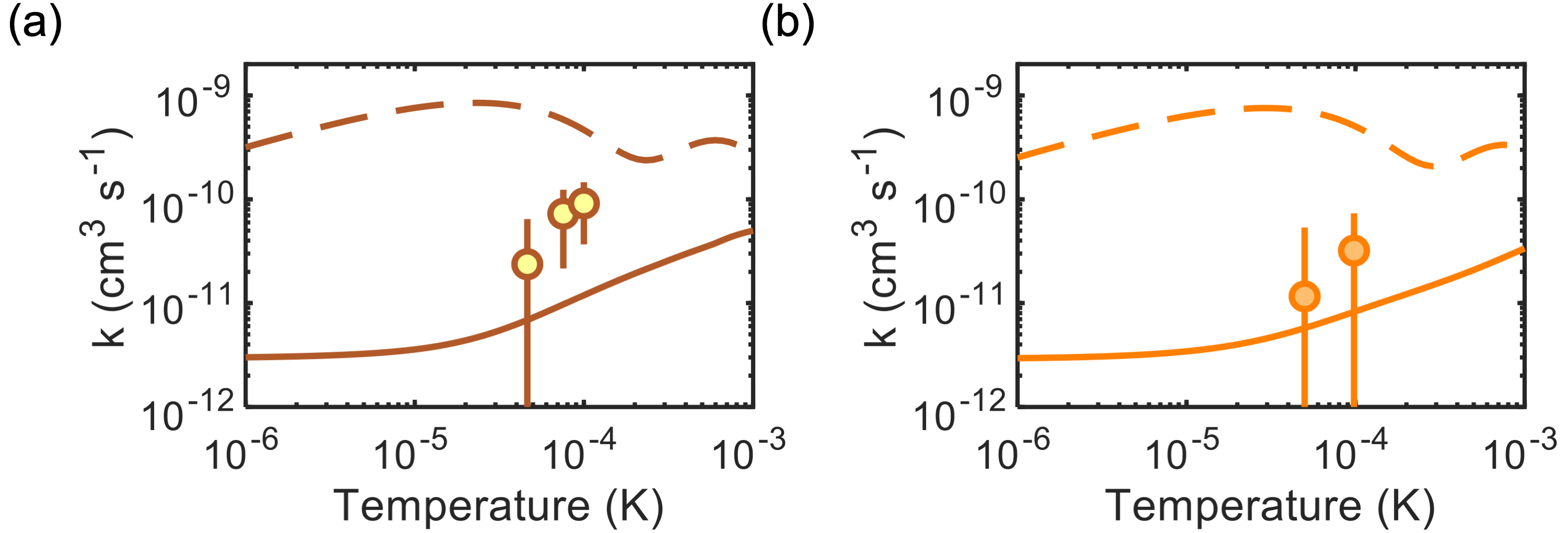}
    \caption{Temperature dependence of rate constants for states with repulsive long-range potentials. Collision rate constants are plotted vs. temperature for molecules prepared in (a) the $f$ state at $\E=0$ and (b) the $d$ state at $\E=500$~V/cm.
    Points are measured collisional loss rate constants with $1\sigma$ error bars. 
    Solid curves are calculated collisional loss rate coefficients, and dashed curves are elastic collision rate coefficients.}
    \label{fig:4}
\end{figure}

The temperature of the molecules in our experiment is well above the dipolar energy scale
$E_d/k_B \sim 1$~$\mu$K~\cite{bohn2009quasi} (but significantly below the parity-doublet energy $2q_\ell/k_B \approx 2$~mK),
suggesting that further cooling may improve the ratio of elastic to inelastic collisions.
Fig.~\ref{fig:4} shows
measured and
calculated rate constants as a function of temperature
for molecules prepared in (a) the $f$ state at $\E = 0$ and (b) the $d$ state at $\E = 500$~V/cm, both of which are shielded from short-range loss but undergo dipolar relaxation at long range.
To vary the molecule temperature in the experiment, we adiabatically lower the ODT trap depth, which cools the molecules and reduces the density. Note that the decreased AC Stark shifts at lower trap depth could also cause small changes in the rate constant. The calculations in Fig.~\ref{fig:4} are carried out at the highest trap depth, $U_0 = k_B \times 4.5$~mK.

In the temperature regime of the experiment, the inelastic collision cross section is constant, leading to a predicted scaling $k \sim T^{1/2}$ of the collisional loss rate constant vs. temperature $T$, in qualitative agreement with our measurements. Meanwhile,
the
calculated
elastic collision rates increase over the same range. At temperatures below around 10~$\mu$K, the collision rates enter the threshold regime, where for identical bosons the inelastic collision rate is constant, and the elastic rate decreases as $k_\text{el} \sim T^{1/2}$. The calculated maximum shielding factor reaches $\gamma = k_\text{el}/k \approx 200$ for both states at temperatures around 10~$\mu$K, which is favorable for evaporative cooling. However, $\gamma$ drops back below 100 at temperatures $\lesssim$1~$\mu$K due to the threshold scaling laws~\cite{sadeghpour2000collisions}.

The maximum value of $\gamma$ depends on intrinsic molecular parameters, primarily the mass, dipole moment, and parity doublet splitting~\cite{bohn2001inelastic}. Higher shielding factors could potentially be obtained for polyatomic molecules with larger parity doublet splittings, which are expected to suppress inelastic loss~\cite{bohn2001inelastic}.

\section{Conclusion}

We have observed and quantified collisions between ultracold CaOH molecules at temperatures near 100~$\mu$K, prepared the molecules in single quantum states within the parity-doubled vibrational bending mode, and measured the dependence of collisional loss rates on the internal state and the magnitude of the applied electric field.
Our measurements are consistent with theoretical calculations that account only for long-range dipolar interactions and for universal short-range loss.
We have identified certain parity-doublet states in the bending mode that have repulsive long-range potentials, which shield the molecules from short-range loss and reduce the measured collisional loss rates. For these states, the remaining collisional loss is dominated by inelastic dipolar relaxation at long range~\cite{bohn2001inelastic}.
The calculations indicate that the ratio of elastic to inelastic collisions in these states is sufficiently high for evaporative cooling of CaOH at temperatures $\lesssim50$~$\mu$K, though improved state preparation would also be required.
Future experiments at even lower temperatures ($\sim$1~$\mu$K) may enable the observation of electrostatic field-linked states supported by parity-doublet structure~\cite{avdeenkov2003linking, augustovicova2019collisions, chen2023field}. These states could be harnessed to form (CaOH)$_2$ dimers~\cite{augustovicova2019collisions,chen2024ultracold}.
Reaching these temperatures remains a challenge for CaOH, but has been previously achieved for laser-cooled diatomic molecules~\cite{cheuk2018lambda, caldwell2019deep, wu2021high}.

Our observations and theory are expected to apply generally to molecules with parity-doublet structure~\cite{avdeenkov2002collisional,avdeenkov2005ultracold,augustovicova2018no}, which exists generically in polyatomic molecules (and is also found in diatomic molecules with degenerate electronic ground states)~\cite{hutzler2020polyatomic}. For parity-doublet states with repulsive potentials, the shielding factor (which determines the feasibility of evaporative cooling) is expected to increase for molecules with larger parity doublet splittings and to also scale with the molecular dipole moment and mass~\cite{bohn2001inelastic},
pointing the way to new molecules.
There are a large number of polyatomic molecules that are amenable to direct laser cooling, including CaOCH$_3$, CaNH$_2$, and their analogues containing other alkaline-earth metals~\cite{kozyryev2016proposal,mitra2020direct,augenbraun2020molecular,frenett2024vibrational}, providing an opportunity to select a molecule with favorable properties for collisional shielding and quantum degeneracy in future work.

\begin{acknowledgments}
We thank Matthew Frye for valuable discussions.
This material is based upon work supported by the Air Force Office of Scientific Research (AFOSR) under award numbers FA9550-22-1-0288, FA9550-24-1-0060, and FA2386-24-1-4070; by the NSF (award numbers PHY-2109995 and PHY-2409404);
and by the ARO. Support is also acknowledged from the U.S. Department of Energy, Office of Science, National Quantum Information Science
Research Centers, Quantum Systems Accelerator. The theoretical work was supported by the Czech Science Foundation (Grant 22-05935S). PR acknowledges support from the NSF GRFP. JLB gratefully acknowledges support from the JILA Physics Frontier Center, NSF award PHY-2317149.
\end{acknowledgments}

\clearpage
\newpage

\onecolumngrid

\begin{center}

\large{\textbf{Supplemental Material}}
\vspace{.5 cm}

\end{center}
\twocolumngrid

\renewcommand{\thefigure}{S\arabic{figure}}

\section{Experimental details}

\subsection{Experimental sequence}

The experiment begins with an RF MOT of CaOH~\cite{vilas2022magneto}, followed by compression, $\Lambda$-cooling, and transfer into a blue-detuned conveyor belt MOT~\cite{hallas2024high, li2025conveyor}. The molecules are transferred into the ODT using 20~ms of single-frequency (SF) cooling~\cite{hallas2023optical}, with the ODT at a trap depth of 4.5~mK. The trap depth is then ramped down to 640~$\mu$K to reduce the molecule density and slow down the collision rate. After a 25~ms hold time to allow untrapped molecules to fall away, the trapped molecules are imaged \emph{in situ} with a 3-5~ms pulse of SF cooling light. This is used to determine the number of molecules initially loaded in the trap.
After the initial SF image (single photon detuning $\Delta \approx 90$~MHz, peak intensity $I_0 \approx 70$~mW/cm$^2$), the molecules are recooled with a 2~ms pulse of SF cooling light (single photon detuning $\Delta \approx 90$~MHz, peak intensity $I_0 \approx 35$~mW/cm$^2$).
We keep the repumping lasers on for 1-2~ms after each imaging and cooling pulse to repump molecules back into the ground vibrational state prior to hold times in the dark. This helps minimize the accumulation of molecules in dark vibrational states~\cite{vilas2023blackbody}.

Next, molecules are prepared in the desired internal state for collision measurements, as described in the sections below. This state preparation is done at trap depths ranging from $\sim$260-960~$\mu$K depending on the specific state (see below). The duration of the state preparation sequence is kept as short as possible to minimize the number of molecules transferred to dark vibrational states via blackbody radiation (BBR) and spontaneous decay~\cite{vilas2023blackbody}. At several points during state preparation, we push detectable molecules that are in the wrong state out of the trap using resonant light (this is referred to as a ``pushout'' pulse below). We estimate that $\sim$80\% of molecules in the trap are in the correct state during the single-state collision measurements, limited by molecules in undetectable vibrational states from optical cycling losses and BBR (see section~\ref{sec:PopulationBackground} below).

After state preparation, we adiabatically ramp the trap depth up to 4.5~mK (over 2~ms) while also ramping the applied electric field to the desired value. (For the temperature scans in Fig.~4 of the main text, we instead ramp up to variable trap depth $\leq$4.5~mK.) We hold the molecules for a variable time at high trap depth, then ramp the trap depth back down to $<1$~mK (and also ramp down the electric field). Next, molecules in detectable states besides the desired collision state are pushed out of the trap, and molecules in the desired collision state are then transferred to a detectable level. This procedure ensures that the only molecules that will be imaged are those that were in the target collision state.
The trap is then turned off, and the molecule cloud is imaged using a 40~ms $\Lambda$-imaging pulse in free space (single-photon detuning $\Delta \approx 12$~MHz, peak intensity $I_0 \approx 70$~mW/cm$^2$). We use free-space imaging to reduce the molecule number density so that light-assisted collisions do not occur during imaging. Such light-assisted collisions could cause systematic errors in the measured collision rates if we were to directly image molecules in the ODT.

\subsection{State preparation in hyperfine mixtures in $\widetilde{X}(000)$ and $\widetilde{X}(010)$}

Here we describe the state preparation for the data in Fig.~1 of the main text.

To prepare molecules in $\widetilde{X}(000)$, we do nothing after SF cooling and imaging, which prepares most molecules in the $(N=1, J=3/2)$ state due to it being the ground state farthest detuned from the cooling light. These molecules are distributed among 8 hyperfine states (magnetic sublevels of $F=1$ and $F=2$, which are spaced by 1.5~MHz). The majority are likely in the $F=2$ manifold, which supports coherent dark states used for deep laser cooling~\cite{caldwell2019deep}.

To prepare molecules in the $\widetilde{X}(010)$ state, we apply a 5~ms pulse of 609~nm $\widetilde{X}^2\Sigma^+(000)(N=1) \rightarrow \widetilde{A}(010)\kappa^2\Sigma^{(-)}(J=1/2^+)$ light (along with all repumping lasers except $\widetilde{X}(010)(N=1)$). The $\widetilde{A}(010)\kappa^2\Sigma^{(-)}(J=1/2^+)$ state decays primarily to $(N=1, J=1/2^-)$ and to $(N=2, J=3/2^-)$, which is repumped back into $(N=1, J=1/2^-)$ via one of the repumping lasers already used for laser cooling. After the bending mode transfer, we push all other detectable molecules out of the ODT.
At this point, molecules are distributed primarily among four hyperfine states (magnetic sublevels of $F=0,1$, which are split by 1.4~MHz) in the $J=1/2^-$ manifold.

\subsection{Preparing single hyperfine states}

We prepare molecules in single hyperfine states for the measurements in Figs.~2 and 4 of the main text as follows. 
We first load the ODT and then ramp the trap depth down to approximately 260~$\mu$K. We then pump the molecules into the bending mode using the 609~nm $\widetilde{X}(000) \rightarrow \widetilde{A}(010)\kappa^2\Sigma^{(-)}$ transition (with repumping lasers),
which prepares most of the molecules in a hyperfine mixture within $\widetilde{X}(010)(N=1, J=1/2^-)$ after approximately 2 photons are scattered. 
Any detectable molecules not successfully pumped into $\widetilde{X}(010)(N=1)$ are pushed out of the trap.

\begin{figure}
\centering
\includegraphics[]{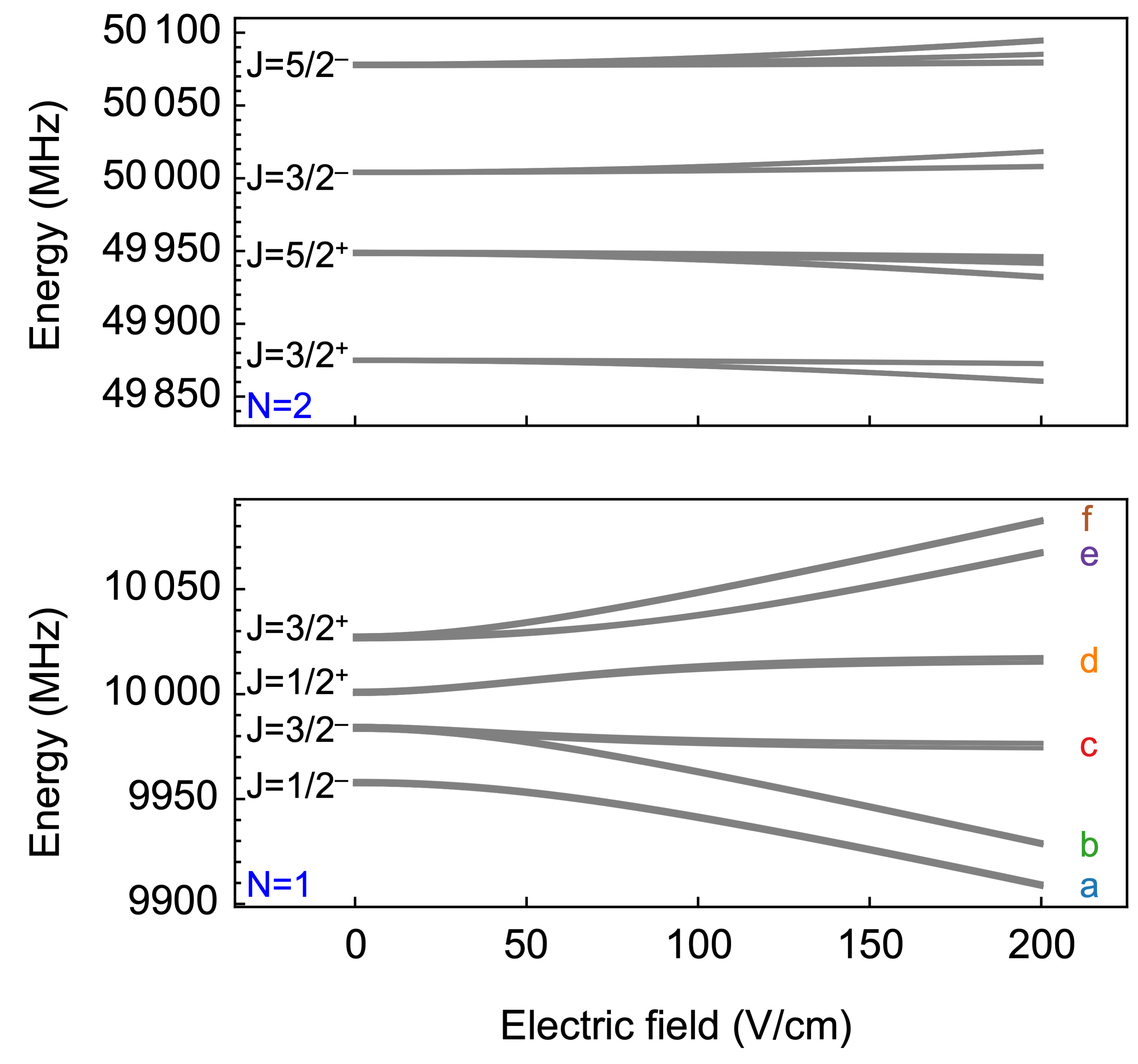}
\caption{Energies of the CaOH $\widetilde{X}(010)$ $N=1$ and $N=2$ manifolds as a function of applied electric field.}
\label{fig:stark_plot}
\end{figure}

Next, molecules are optically pumped into the $(J=1/2^-, F=0)$ hyperfine state using a combination of microwave and optical frequencies as described in Ref.~\cite{anderegg2023quantum}, though in this work the $J=3/2^-$ spin-rotation component is not addressed. The optical pumping efficiency is $\sim70\%$. Molecules successfully pumped into $(J=1/2^-, F=0)$ are then transferred into the desired collision state using microwave and RF pulses to drive transitions within and between the two rotational states $N=1$ and $N=2$ in $\widetilde{X}(010)$. The structure of these rotational states is shown in Fig.~\ref{fig:stark_plot}.

\begin{table}
    \centering
    \begin{tabular}{c|c|c}
    \hline \hline
        Manifold & B (G) & Hyperfine state \\
        \hline
        $a$ & 1.8 & $(J=1/2^-, F=0, m_F=0)$\\
        $b$ & 2.9 & $(J=3/2^-, F=2, m_F=2)$\\
        $c$ & 2.9 & $(J=3/2^-, F=2, m_F=1)$ \\
        $d$ & 0 & $(J=1/2^+, F=0, m_F=0)$\\
        $e$ & 2.0 & $(J=3/2^+,F=2,m_F=0)$\\
        $f$ & 1.8 & $(J=3/2^+, F=2, m_F = 2)$ \\
        \hline \hline
    \end{tabular}
    \caption{Magnetic field and hyperfine state used for each collision manifold in the experiment.}
    \label{tab:CollisionMagneticFields}
\end{table}

The six hyperfine states studied in this work are listed in Tab.~\ref{tab:CollisionMagneticFields}, along with the magnetic field applied so that the state is resolved across the range of electric fields studied.
The details of the state preparation are specific to each collision state and are chosen to ensure that the state does not undergo level crossings as the trap depth and fields are ramped. In all our state preparation sequences, we use pushout pulses to drive away population that has accumulated in detectable impurity states as a result of state preparation inefficiencies, radiative decay, or blackbody radiation. By including or excluding repumping lasers for $N=1$ and $N=2$ during the pushout, we can selectively drive away population from these rotational states. During these pushouts, molecules can be shelved in undetectable states (usually states of positive parity within $N=1$ or $N=2$; the repumping lasers only address negative parity) that are not pushed out. In the following sections, we describe the preparation of each of the collision states $a$--$f$.

\subsubsection{Preparing the a and f states}

To prepare the $a$ and $f$ states, we drive RF $\pi$ pulses between the $(N=1, J=1/2^-, F=0) \leftrightarrow (N=1, J=3/2^+, F=2, m_F=2)$ states at $E=0$~V/cm and $B = 1.8$~G. Because this transition is only allowed by tensor AC Stark shifts (which mix states with $\Delta M_F = \pm 2$), we use a relatively high trap depth of $\sim$960~$\mu$K when driving this transition. The typical $\pi$ pulse efficiency is around 90\%. For the $a$ state, we drive a $\pi$ pulse up to $f$ (which is dark to the detection lasers), apply a pushout, and finally drive a second $\pi$ pulse back down to $a$. For the $f$ state, we apply a $\pi$ pulse up and push out, but then leave the molecules in the upper state. After the collision hold time, we reverse the state preparation, being sure to apply another pushout
before detecting.

\subsubsection{Preparing the b and c states}

The $b$ and $c$ states both start as $J=3/2^-$ at $\mathcal{E}=0$, then split apart at fields above $\mathcal{E}\sim60$~V/cm.
To reach both states, we use two 40~GHz microwave transitions via $N=2$, since direct RF transitions from $a$ to $b$ and $c$ are electric dipole forbidden. We begin by turning on a 2.9~G magnetic field, then transfer population from $(N=1, J=1/2^-, F=0) \rightarrow (N=2, J=3/2^+, F=2, m_F=1)$\footnote{Hyperfine states are significantly mixed at this magnetic field, so $F$ is not a good quantum number. We label these states by their dominant $F$ component at 2.9~G, but in reality the ${(N=2, F=2, m_F=1)}$ state correlates to ${(N=2, F=1, m_F=1)}$ at zero field.} via adiabatic rapid passage (ARP), by adiabatically sweeping the electric field across resonance in the presence of microwave radiation. Next, we apply a pushout pulse, noting that molecules in $(N=2, J=3/2^+)$ are dark to the detection lasers and thus remain in the trap.

For the $\mathcal{E}=0$ data, we next use another ARP electric field sweep to transfer population down to $(N=1, J=3/2^-, F=2, m_F=2)$. For data at fields $\gtrsim60$~V/cm, we ramp the electric field to approximately 90~V/cm with population still in $N=2$, and then use another ARP electric field sweep to transfer population down to either the $b$ or the $c$ state.
After the collision hold, we push out all detectable molecules except those in $\widetilde{X}(010)(N=1)$, then detect.

\subsubsection{Preparing the d state}

To prepare molecules in the $d$ state, we again use two microwave transitions via $N=2$.
In this case, we start by applying a 2~G magnetic field, then ramp the electric field up to 150~V/cm and use an ARP electric field sweep to transfer population to $(N=2, J=3/2^-, F=1, m_F=-1)$.
Next, we ramp the electric field down to $\mathcal{E}=50$~V/cm, then ramp the magnetic field down to 0~G. The fields are ramped in this order since the $m_F=\pm 1$ states are nearly degenerate at 150~V/cm, but not at 50~V/cm.
Finally, we drive population down to the $d$ state $(N=1, J=1/2^+, F=0)$ using an ARP electric field sweep.
At 0~G magnetic field, the $d$ state undergoes no level crossings between $\mathcal{E}=0-600$~V/cm.
Both before and after the collision hold, we ramp the electric field to $\mathcal{E}=0$ and apply a pushout pulse. To detect molecules at the end of the sequence, we transfer molecules back to $N=2, J=3/2^-$, where they can be imaged.

\subsubsection{Preparing the e state}

The $e$ state undergoes significant avoided crossings below around 90~V/cm, and at zero field is the same as the $f$ state  (Fig.~\ref{fig:stark_plot}). Therefore, we only prepare the $e$ state at fields above 90~V/cm. We start the same way as for the $d$ state, using an ARP electric field sweep to populate $(N=2, J=3/2^-, F=1, m_F=-1)$ at $E=150$~V/cm and $B=2$~G. Next, we ramp the electric field down to $\mathcal{E}=0$ and apply a pushout pulse for all detectable states except $\widetilde{X}(010)(N=2)$, where the molecules are shelved. The electric field is then ramped back up to approximately 100~V/cm, and another ARP electric field sweep is used to drive down to the $e$ state.
After the collision sequence, 
we 
use an ARP electric field sweep to drive population back up to $N=2$, ramp down the electric field, and push out all detectable molecules except those in $\widetilde{X}(010)(N=2)$. Finally, we repump the molecules in $N=2$ and detect.

\subsection{Single-molecule lifetimes}

We verified that the (blackbody-limited) single-molecule lifetime for most states is consistent with previous measurements~\cite{hallas2023optical, vilas2023blackbody}, even at the full 4.5~mK ODT trap depth. The previously measured values (which we also use here) are $\tau = 900$~ms for $\widetilde{X}(000)$ and $\tau = 360$~ms for the $\widetilde{X}(010)$ bending mode. The exception is the $f$ state, which has a measured $\sim250$~ms lifetime at full trap depth, increasing to $\sim360$~ms as the trap is ramped down. This is accounted for when fitting the collision rate constants.

\section{Calibration of molecule density}

\subsection{Calibration of molecule number}

The molecule numbers were calibrated as follows. The collection efficiency of the imaging system (based on a $\sim$0.4~NA in-vacuum aspheric lens) was estimated to be $C=0.55(12)$\% for molecules trapped in the ODT. This was done by measuring the signal collected when imaging single molecules in the optical trap, and also comparing to measurements of the collection efficiency from previous experiments utilizing a very similar imaging system~\cite{vilas2024optical, hallas2024high}.

For the collision data, the molecules were imaged in a free-space molasses after turning off the trap, changing the collection efficiency compared to trapped molecules due to the larger size of the cloud and nonuniform imaging beam intensities. To account for this effect, the signal obtained from \emph{in situ} imaging of a small number of molecules in the optical trap (with a known scattering rate) was compared to the signal obtained from imaging the same number of molecules in free space. The combined fractional uncertainty in the molecule number, including estimated errors in the collection efficiency, \emph{in situ} imaging scattering rate, and free-space imaging conversion factor, is approximately 25\%.

\subsection{Molecule temperature}

The molecule temperature was measured using time-of-flight (TOF) expansion. After loading the molecules into the ODT, an SF cooling pulse (single-photon detuning $\Delta\approx 90$~MHz, peak intensity $I_0 \approx 35$~mW/cm$^2$) was applied to set the initial temperature. 
The ODT intensity was then ramped up to the collision trap depth ($U_0 \approx k_B \times 4.5$~mK), held there for 1 ms, and then instantaneously shut off. The molecules were allowed to expand for a variable amount of time, before being imaged with a 10~ms pulse of $\Lambda$-cooling light using the large, 10~mm~$1/e^2$-diameter MOT beams.

For molecules in hyperfine mixtures (Fig.~1 of the main text), no state preparation was done after the SF cooling pulse, and the temperature was measured to be 80~$\mu$K.
For molecules in single hyperfine states in $\widetilde{X}(010)$ (Fig.~2 of the main text), after the SF cooling pulse the molecules were optically pumped into the bending mode and then microwave-optically pumped into the $(J=1/2^-, F=0)$ hyperfine state. The optical pumping causes some heating, and the temperature of these molecules was measured to be 100~$\mu$K during the collision time.

For the temperature scans in Fig.~4 of the main text, we adiabatically ramp the trap depth to different values between $U_0 = k_B \times 960~\mu$K and $U_0 = k_B \times 4.5$~mK. The temperature scales as $T\propto \sqrt{U_0}$ when adiabatically ramping the trap depth, which we confirmed using TOF temperature measurements.

\subsection{Optical trap geometry}

The ODT trap frequencies were measured using parametric heating spectroscopy, whereby the trap laser intensity was modulated to induce heating, and the resulting molecule loss was recorded as a function of the modulation frequency. Three loss features were observed, corresponding to approximately $2\times$ the motional frequency along each principal axis of the trap. These measurements were taken at relatively low trap depth to minimize the influence of collisions on the measurement, then scaled up using $\omega_i \propto \sqrt{U_0} \propto \sqrt{P}$, where $\omega_i$ is the trap frequency along axis $i$, $U_0$ is the trap depth, and $P$ is the laser power (this scaling was experimentally verified).
At the trap depth used for the collision measurements ($P=12.6$~W), the scaled trap frequencies were $\{\omega_x^\text{meas}, \omega_y^\text{meas}, \omega_z^\text{meas}\} = 2\pi \times \{30.4, 26.2, 0.45 \}$~kHz. However, Monte Carlo simulations of the parametric heating procedure (using the Gaussian trap geometry described below) indicate that the molecules' finite temperature causes our measurement to underestimate the axial trap frequency, $\omega_z$, by $\sim$20\%. Accounting for this effect gives ``corrected'' frequencies $\{\omega_x, \omega_y, \omega_z\} = 2\pi \times \{30.4, 26.2, 0.55 \}$~kHz.

Next, we infer the trap geometry from the measured frequencies.
The trapping laser is modeled as an elliptical Gaussian beam, whose waists ($w_x, w_y$), Rayleigh range ($z_R$), and trap depth ($U_0$) are related to the trap frequencies and laser power by
\begin{align}
\omega_x &= \sqrt{\frac{4U_0}{mw_x^2}}, \quad \omega_y = \sqrt{\frac{4U_0}{mw_y^2}}, \quad \omega_z = \sqrt{\frac{2U_0}{m z_R^2}}, \nonumber \\
U_0 &= \frac{1}{2\epsilon_0 c}\alpha_0 I_0,
\label{eqn:trapfrequencies}
\end{align}
where $I_0=2P/(\pi w_x w_y)$ is the peak laser intensity and $\alpha_0 = 204$~a.u. is the scalar polarizability of CaOH at 1064~nm~\cite{hallas2023optical}.
Using the corrected frequencies from above, the trap parameters are estimated to be $w_x = 8.4$~$\mu$m, $w_y = 9.8$~$\mu$m, and $z_R = 330$~$\mu$m, implying a trap depth of $U_0/k_B = 4.5$~mK at the trap power used for the collision data. Note that the modeled Rayleigh range, $z_R$, is slightly larger than the idealized Rayleigh ranges ($z_{R,x} = \pi w_x^2/\lambda \approx 210$~$\mu$m, $z_{R,y} = \pi w_y^2/\lambda \approx 280$~$\mu$m) that would correspond to the modeled beam waists. This could be due to aberrations (e.g. astigmatism) that are not captured in the model. Such aberrations should not significantly affect the final density determination since $\eta = U_0/k_B T \approx 45$, so the molecules see only the very bottom of the trap.

\subsection{Number density}

The average number density of molecules in the ODT is
\begin{equation}
\langle n\rangle = \frac{N \omega_x \omega_y \omega_z}{\xi(T)(4\pi k_B T/m)^{3/2}} \equiv \frac{N}{V_\text{eff}(T)}
\end{equation}
where $N$ is the molecule number, $V_\text{eff}(T) = \xi(T)(4\pi k_B T/m)^{3/2}(\omega_x \omega_y \omega_z)^{-1}$ is the effective trap volume, and $\xi(T)$ is a correction factor accounting for deviations from the harmonic approximation at finite temperature ($\xi = 1$ for a purely harmonic trap). $\xi(T)$ is just the ratio between the average number densities calculated in the harmonic approximation and for the actual trap geometry, and can be computed by integrating the number density over space. For the Gaussian trap model described above, we calculate $\xi(T = 80~\mu\text{K}) \approx 1.07$ and $\xi(T=100~\mu\text{K}) \approx 1.09$.

Using the Gaussian model, for a temperature of $T=100~\mu$K and $N=100$ molecules in the trap, we calculate a peak number density $n_0 \approx 3.7 \times 10^{11}$~cm$^{-3}$, an average number density $\langle n \rangle \approx 1.3 \times 10^{11}$~cm$^{-3}$, and a peak phase space density $\rho_0 = n_0 \lambda_\text{dB}^3 \approx 5 \times 10^{-6}$, where $\lambda_\text{dB} = h/\sqrt{2\pi m k_B T}$ is the thermal de Broglie wavelength.

The dominant uncertainty in the number density is the $\sim$25\% fractional uncertainty in the molecule number calibration. The uncertainties in the trap volume and molecule temperature are comparatively small. Therefore, we estimate that the total fractional uncertainty (standard error) in the molecule number density is 25\%.

\section{Effect of population impurities}
\label{sec:PopulationBackground}

\subsection{``Background'' collision rates}
\label{sec:BackgroundRates}

Suppose the molecules are imperfectly prepared in the target hyperfine state, such that the total number density $n$ in the trap is the sum of the density of molecules in the target state, $n_1$, and that of molecules in impurity states, $n_2$. The rate equation describing two-body collisions of molecules in the target state is
\begin{align}
\dot{n}_1 &= -k_{11} n_1^2 - k_{12}n_2n_1 = -\left(k_{11}+k_{12}\frac{n_2}{n_1}\right)n_1^2 \nonumber \\
&\equiv -\left[k_{11} + k_\text{bg}(n_1,n_2)\right]n_1^2 \nonumber \\
&\equiv -k_\text{eff}(n_1,n_2)n_1^2
\end{align}
where $k_{11}$ is the loss rate constant for collisions between two molecules in the target state and $k_{12}$ is the rate constant for collisions between a target molecule and an impurity molecule. We have rewritten the equation for $\dot{n}_1$ in terms of an effective collision rate, $k_\text{eff} = k_{11} + k_\text{bg}$, which is the effective, instantaneous two-body collision rate for molecules in state 1. In the experiment, we wish to measure $k_{11}$ but have to contend with background collisions due to impurity molecules, described by the ``background'' rate constant $k_\text{bg}$.
Even if collisions between molecules in the target state are fully suppressed ($k_{11}=0$), we will measure a nonzero collision rate $k_\text{bg} = k_{12} n_2/n_1$.
Generalizing to the case where there are many impurity states $i$, the background loss rate is a sum over all impurity states, $k_\text{bg} = \sum_i k_{1i} n_i/n_1$.

Molecules end up in background, impurity states due to rovibrational state redistribution from blackbody radiation and spontaneous decay, and from losses to vibrational dark states during optical cycling. These effects are modeled using rate equations, as described below. For the single-state collision measurements in Fig.~2 of the main text, we estimate that $78(8)\%$ of molecules are in the target state during the collision hold time, and that the average collision rate between target molecules and impurity molecules is $\langle k_{1i} \rangle \approx 2.5(5) \times 10^{-10}$~cm$^3$~s$^{-1}$. This gives a background collision rate $k_\text{bg} \approx 7(3) \times 10^{-11}$~cm$^3$~s$^{-1}$.
This same background rate constant is also used for the temperature scans in Fig.~4 of the main text, since the background rate constants are assumed to be near the Langevin capture limit, which depends only weakly on temperature~\cite{jurgilas_2021, bell2009ultracold}.

For the hyperfine mixture data in Fig.~1 of the main text, the state preparation is better since no optical pumping into a single hyperfine state is required and overall hold times are shorter. For these measurements, the rate equation model indicates that the background collision rates are $k_{\text{bg},(000)} \approx 0.5(5) \times 10^{-11}$~cm$^3$~s$^{-1}$ and $k_{\text{bg},(010)} \approx 3(2) \times 10^{-11}$~cm$^3$~s$^{-1}$ for $\widetilde{X}(000)$ and $\widetilde{X}(010)$, respectively.

\subsection{Accounting for background collisions in measured rate constants}

To account for background collisions, we subtract off the estimated $k_\text{bg}$ from the measured rate and account for the corresponding error. In detail, we first fit the rate constant from the experimental data using eqn.~1 in the main text, which has a mean value $k_\text{meas}$ and statistical uncertainty (standard error) $\sigma_\text{meas}$. We then subtract off the estimated background collision rate to obtain a mean and standard error for the ``corrected'' collision rate:
\begin{align*}
k &= k_\text{meas} - k_\text{bg} \\
\sigma &\approx \sqrt{\sigma_\text{meas}^2 + (k_\text{meas}\sigma_{n})^2 + \sigma_\text{bg}^2}
\end{align*}
where $\sigma_n = 0.25$ is the error in the density calibration and $\sigma_\text{bg}$ is the error in $k_\text{bg}$ (described above).

\subsection{Rate equation modeling of background collisions}

We model the effect of trapped CaOH molecules in dark states using a rate equation model similar to the one described in Ref.~\cite{vilas2023blackbody}. The model includes (1) rovibrational state distribution driven by blackbody radiation (BBR) and spontaneous decay, (2) loss of molecules to vibrational dark states during optical cycling, and (3) lossy two-body collisions between trapped molecules in all states.
Mechanism (1) is modeled as described in Ref.~\cite{vilas2023blackbody}, while mechanisms (2) and (3) have been added in this work.
We include vibrational states up to $v_1=2$, $v_2=2$, and $v_3=0$, and include rotational states up to $N=5$ in each vibrational manifold (fine structure states $J$ are also included, but hyperfine structure is neglected).

To model optical cycling losses, molecules are transferred to the $\widetilde{X}(22^00)(N=1, J=3/2)$ state with a probability $P_\text{loss} \approx 7 \times 10^{-5}$ per scattered photon during ODT loading and imaging.\footnote{Optical cycling losses in CaOH are likely dominated by decay to the $(220)$, $(040)$, and $(030)(N=2)$ states~\cite{zhang2021accurate, vilas2022magneto}. However, here we approximate all losses as ending up in $\widetilde{X}(220)$ since it is the only one of these states included in the model. The key point is that the losses are to a relatively high-lying state in the vibrational potential, which can quickly (over $\sim$100~ms) decay down to a number of lower-lying states (some of which may even be detectable).} Light-assisted collisions during loading and imaging are also included, with a measured rate constant $k_\text{SF} \approx 2 \times 10^{-10}$~cm$^3$~s$^{-1}$ for SF cooling of molecules in bright states.

\begin{figure*}
\centering
\includegraphics{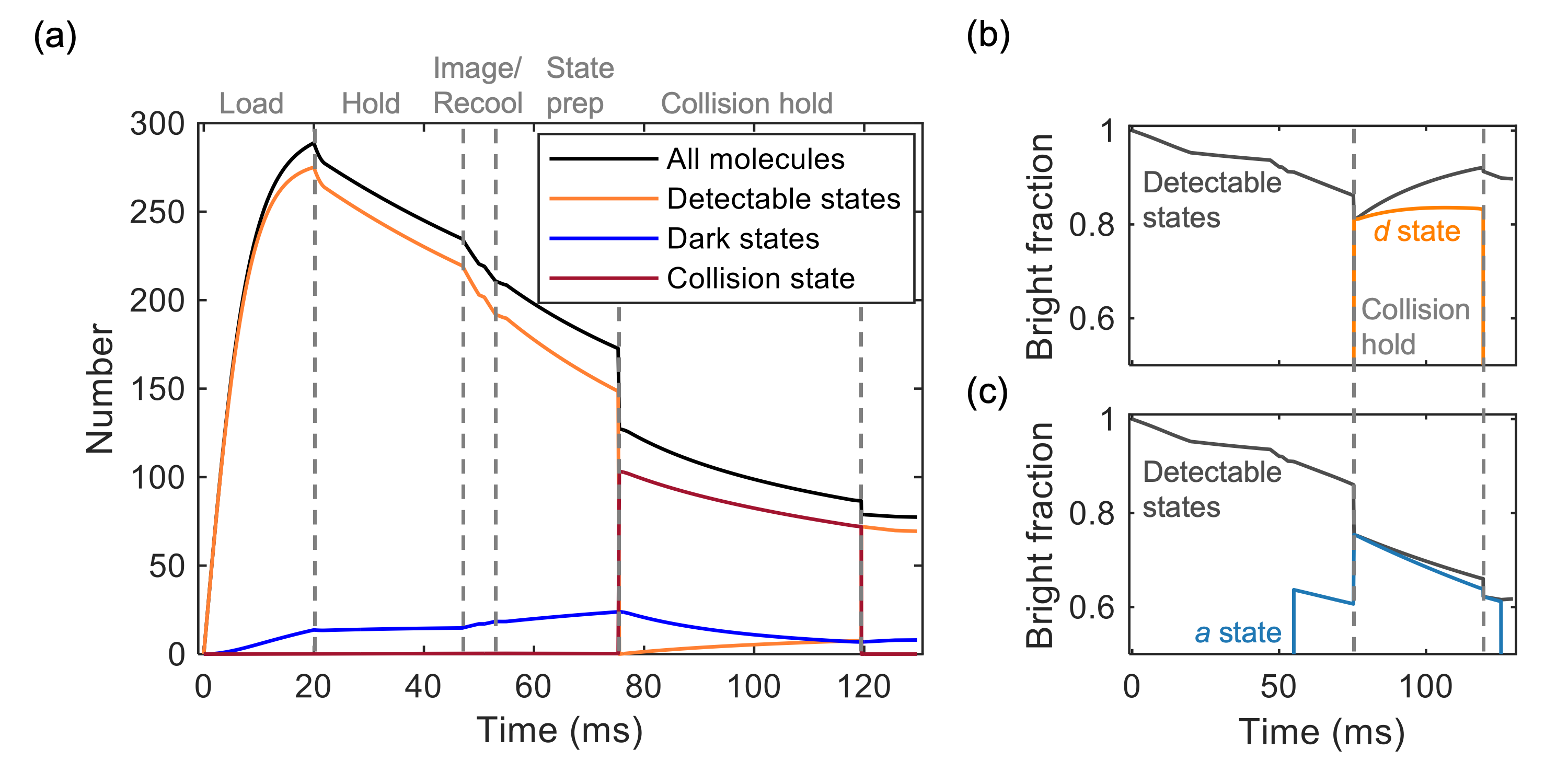}
\caption{Rate equation model of dark-state population during the experimental sequence. (a) Modeled state populations during the sequence used for collisions in the $d$ state at $\mathcal{E}=0$. The total number of trapped molecules, the molecules in optically detectable states, the number in undetected (dark) states, and the number in the target collision state ($d$ in this case) are plotted as a function of time after trap loading. (b) Fraction of molecules in detectable states and in the $d$ state, for the $d$ state collision sequence at $\mathcal{E}=0$. (c) Fraction of molecules in bright states and in the $a$ state during the $a$ state collision sequence at $\mathcal{E}=0$.} 
\label{fig:RateEquations}
\end{figure*}

\begin{figure*}
\centering
\includegraphics{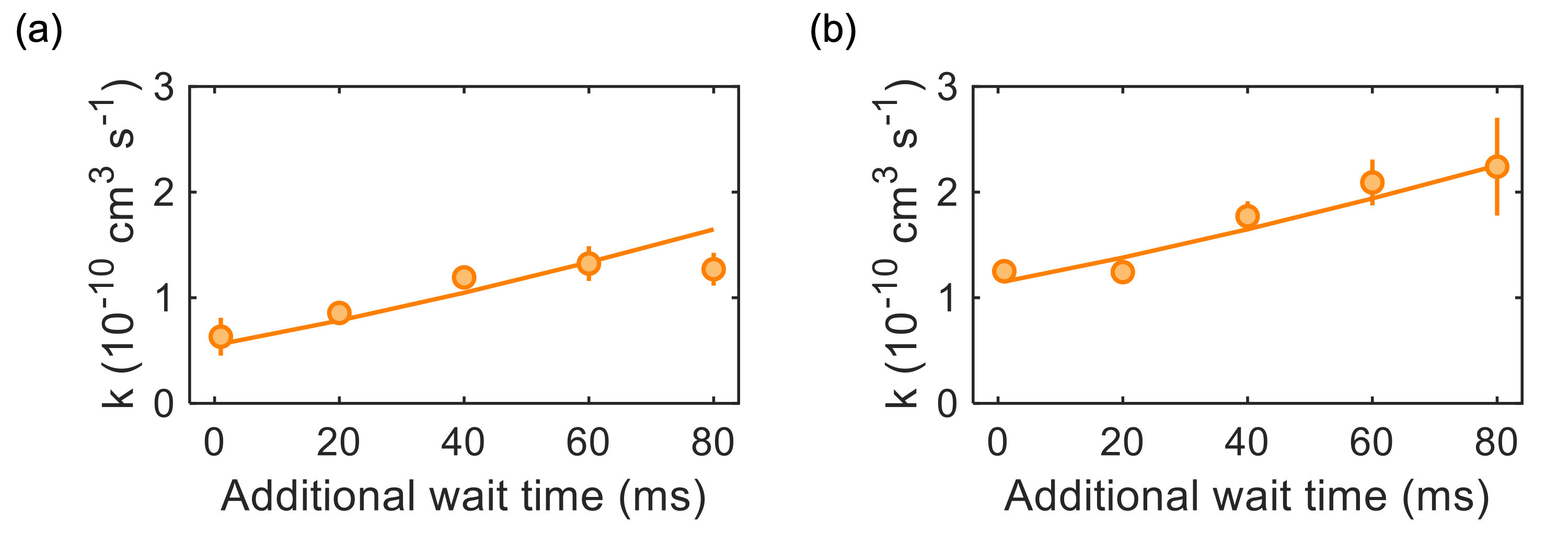}
\caption{Fitted collisional loss rate constant vs. wait time (at low density) prior to the collision hold for (a) the $d$ state at $\mathcal{E}=0$ and (b) the $d$ state at $\mathcal{E}=500$~V/cm. The fitted rate constant increases as a function of hold time since more molecules end up in background states, increasing $k_\text{bg}$. Solid curves are calculated from the rate equation model described in the text, which accounts for all CaOH molecules trapped in the ODT, including in dark rovibrational states. The collision data in the main text were taken at $t=0$, which is
the minimum hold time needed for state preparation.} 
\label{fig:HoldTime}
\end{figure*}

To model two-body collisions, we use a matrix of loss rate constants $k_{ij}$ that describes collisions between pairs of rovibrational states $i$ and $j$ included in the model. 
The loss rate constants $k_{ij}$ are important for determining the background loss rate, $k_\text{bg}$, but are unknown for most combinations of states $i$ and $j$. We estimate these rates using calculated universal/Langevin rate constants along with measurements taken during this work. The state pairs fall into two categories. The first consists of states $i$ and $j$ coupled by an electric dipole-allowed transition, which could be a vibrational transition (selection rules $\Delta v_i = \pm 1, \Delta v_{j\neq i} = 0, \Delta N = 0, \pm 1, p = +1 \leftrightarrow -1$)~\cite{vilas2023blackbody} or a rotational transition. These rate constants are estimated using the Langevin capture model for a $-C_3/r^3$ potential~\cite{bell2009ultracold}, with $C_3 = |\langle i | d | j \rangle|^2/(4\pi \epsilon_0)$, where $d$ is the electric dipole operator. They typically fall in the range $k_{ij} \sim 1-10 \times 10^{-10}$~cm$^3$~s$^{-1}$.
The second category consists of pairs of states which are not directly coupled by an electric dipole transition. These states generally interact via attractive $-C_6/r^6$ (van der Waals, vdW) interactions, whereby the electric dipole interaction acts at second order (see section~\ref{sec:DDI} below). For two molecules in any $\ell=0$ state, we use the measured rate constant $k = 0.7 \times 10^{-10}$~cm$^3$~s$^{-1}$ for molecules in $\widetilde{X}(000)$, since in all cases the dominant vdW interaction comes from mixing of rotational states. For one molecule in an $\ell = 0$ state and one molecule in an $\ell = 1$ state, $C_6$ is typically higher since the second-order dipolar coupling mixes rotational states in the $\ell = 0$ molecule but only needs to mix parity-doublet states in the $\ell = 1$ molecule. Thus, the energy denominator in the second-order perturbation theory sum (eqn.~\ref{eqn:vdWsum}) is generally smaller than in the $\ell = 0 + \ell = 0$ case. We estimate these collision rates using the Langevin capture model described in sec.~\ref{sec:Universal} and find rates ranging from $k_{ij}(\ell_i = 0, \ell_j = 1) \sim 1-5 \times 10^{-10}$~cm$^3$~s$^{-1}$ depending on the combination of rotational states (an ``effective'' rate for $\ell = 0 + \ell = 1$ collisions is fit as described below). Finally, for two molecules in $\ell \neq 0$ states, vdW interactions arise from mixing of closely-spaced parity doublet states, and we use $k_{ij}(\ell_i \neq 0, \ell_j \neq 0) \approx 5 \times 10^{-10}$~cm$^3$~s$^{-1}$, which is similar to the universal rate for two molecules in $\widetilde{X}(010)(N=1^-)$. Note that in all cases these rate constants are at best semi-quantitative estimates, and in reality there will be significant dependence on the specific rotational state. However, they do provide a starting point for the rate equation model and can be expected to give reasonable results when averaged over the large number of states in the model.

We use the rate equation model to simulate the entire experimental sequence, including ODT loading, imaging, state preparation, and trap depth ramps. The simulated molecular populations as a function of time for the $d$ state collision measurement sequence are plotted in Fig.~\ref{fig:RateEquations}(a). The state preparation pulses are approximated as instantaneous, with a microwave-optical pumping efficiency of 70\%, and coherent state transfer efficiencies of 90-100\%.

To benchmark the model, we took data with molecules held for a variable time in $\widetilde{X}(010)(N=1,J=1/2^-)$ after optical pumping and before the collision hold, in order to intentionally increase the number of molecules in impurity states. This was done for the $d$ state at both $\mathcal{E}=0$ and $\mathcal{E}=500$~V/cm. Fig.~\ref{fig:HoldTime} shows the fitted rate constants as a function of this additional hold time, along with the results of the rate equation model. Because the primary effect of the hold time is to allow molecules in $\widetilde{X}(010)(N=1,J=1/2^-)$ to decay to $\widetilde{X}(000)(N=0,2)$, the slope of these curves is primarily sensitive to the rate constants $k_{ij}(\ell_i = 0, \ell_j = 1)$ between $\ell = 0$ and $\ell = 1$ states of the same parity. We find that the rate equations match the data well if this rate is set to $k_{ij}(\ell_i = 0, \ell_j = 1) = 2.5 \times 10^{-10}$~cm$^3$~s$^{-1}$.

Fig.~\ref{fig:RateEquations}(b-c) shows the fraction of molecules in the target collision state during the collision sequence for (b) the $d$ state at $\mathcal{E}=0$ and (c) the $a$ state at $\mathcal{E}=0$, as modeled by the rate equations.
In both cases, we see that approximately 75-80\% of molecules are prepared in the target state at the start of the collision hold time.
By varying experimental parameters in the model, we estimate that the population in the target state can vary over the range $\sim70-85$\%, corresponding to an estimated mean and uncertainty in the state preparation efficiency of 78(8)\%.
The model also allows us to estimate the weighted-average loss rate between target molecules and background molecules, $\langle k_{1i} \rangle = \sum_j k_{1j} n_j / \sum_j n_j$. Over the range of states studied for collisions in the $\widetilde{X}(010)$ bending mode, we estimate $\langle k_{1i} \rangle = 2.5(5) \times 10^{-10}$~cm$^3$~s$^{-1}$. This is used to estimate $k_\text{bg}$, as described in sec.~\ref{sec:BackgroundRates}.

\section{Dipolar Interaction Potentials}
\label{sec:DDI}

\subsection{Single-molecule Hamiltonian}

The effective Hamiltonian for CaOH in the $\widetilde{X}(010)$ vibrational bending mode is~\cite{anderegg2023quantum, Hirota1985, merer1971rotational, caldwell2020sideband}:
\begin{equation}
    H_X = H_\text{rot} + H_\text{sr} + H_{\ell \text{d}} + H_\text{hf} + H_\text{S} + H_\text{Z} + H_\text{ac} \label{eqn:heff}
\end{equation}
where the individual terms are
\begin{subequations}
\begin{align}
    &H_\text{rot} = B(\vec{N}^2 - \ell^2) \quad \\
    &H_\text{sr} = \gamma(N_xS_x + N_y S_y) \\
    &H_{\ell \text{d}} = \frac{q_\ell}{2}(N_+^2 + N_-^2)
    \label{eqn:ld}\\
    &H_\text{hf} = b_F \vec{I} \cdot \vec{S} + \frac{c}{3} (3I_zS_z - \vec{I} \cdot \vec{S}) \\
    &H_\text{S} = -\vec{d} \cdot \vec{\mathcal{E}} \\
    &H_\text{Z} = g_S \mu_B \vec{S} \cdot \vec{B} \\
    &H_\text{ac} = -\vec{d} \cdot \vec{\mathcal{E}}_\text{ODT}
\end{align}
\end{subequations}
$H_\text{rot}$ is the molecular rotation with rotational constant $B = 9997$~MHz; $H_\text{sr}$ is the spin-rotation interaction with spin-rotation parameter $\gamma=35.5$~MHz; $H_{\ell \text{d}}$ is the $\ell$-type doubling interaction with $\ell$-doubling parameter $q_\ell=21.5$~MHz; $H_\text{hf}$ is the hyperfine interaction with Fermi contact and dipolar parameters $b_F=2.45$~MHz and $c=2.6$~MHz; $H_\text{S}$ is the DC Stark shift, where $\vec{d}$ is the molecule-frame electric dipole moment with magnitude $|d| = 1.465$~D; $H_\text{Z}$ is the electron-spin Zeeman shift, where $g_S$ is the electron $g$ factor and $\mu_B$ is the Bohr magneton; and $H_\text{ac}$ is the AC Stark shift from the trapping ODT laser, where $\vec{d}$ is the electric dipole operator and $\vec{\mathcal{E}}_\text{ODT}$ is the electric field of the ODT laser. In the experiment, the static electric and magnetic fields $\vec{\mathcal{E}}$ and $\vec{B}$ are aligned along the laboratory $Z$ axis, while the ODT is linearly polarized along the laboratory $X$ axis. Values of the Hamiltonian parameters are taken from Refs.~\cite{li1995bending, steimle1992supersonic, hallas2023optical, anderegg2023quantum}.

Matrix elements of eqn.~\ref{eqn:heff} are evaluated in the Hund's case (b) basis $|N\ell SJIFM\rangle$ using expressions from Ref.~\cite{Hirota1985} for all but the $\ell$-doubling and AC Stark terms. The AC Stark matrix elements are evaluated according to Ref.~\cite{caldwell2020sideband}, using the 1064~nm polarizabilities from Ref.~\cite{hallas2023optical}. The $\ell$-doubling matrix elements are
\begin{align}
\langle N, &\ell, S, J, I, F, M| H_{\ell \text{d}} | N, \ell', S, J, I, F, M \rangle \nonumber \\
&= q_\ell \sum_{q=\pm 1} (-1)^{N-\ell}
\begin{pmatrix}
N & 2 & N \\
-\ell & 2q & \ell'
\end{pmatrix}
\nonumber \\
& \times \frac{1}{2\sqrt{6}} \left\{ (2N+3)(2N+2)(2N+1)(2N)(2N-1)\right\}^{1/2}
\end{align}
which are similar to $\Lambda$-doubling matrix elements from, e.g., Ref.~\cite{brown2003rotational}.

\subsection{Dipolar interaction Hamiltonian}

To describe dipolar interactions between two molecules, we expand our basis to include the internal states of two molecules, plus angular momentum quantum numbers $|L, M_L\rangle$ describing the two-molecule system. The complete two-molecule basis states are therefore $|N_1 \ell_1 S_1 J_1 I_1 F_1 M_1 \rangle |N_2 \ell_2 S_2 J_2 I_2 F_2 M_2\rangle |L M_L\rangle \equiv |\eta L M_L \rangle$, where $|\eta\rangle$ describes the internal quantum numbers of the two molecules.

The internal energies of the two molecules are given by the Hamiltonian
\begin{equation}
H_\text{mol} = H_{X,1} \otimes \mathds{1}_2 + \mathds{1}_1 \otimes H_{X,2}
\end{equation}
The dominant molecule-molecule interaction is the electric dipole-dipole interaction,
\begin{equation}
H_\text{ddi} = \frac{1}{4\pi \epsilon_0 r^3} \left[ \vec{d}_1 \cdot \vec{d}_2 - 3(\hat{r}_{12} \cdot \vec{d}_1 )(\hat{r}_{12} \cdot \vec{d}_2) \right]
\end{equation}
where $\hat{r}_{12}$ is the unit vector pointing from molecule 1 to molecule 2 and $r$ is the separation between the two molecules.

Matrix elements of $H_\text{ddi}$
are~\cite{augustovicova2019collisions}
\begin{widetext}
\begin{align}
\langle \eta L M_L | H_\text{ddi} | \eta' L' M_L' \rangle =&  -\frac{\sqrt{30}}{4\pi\epsilon_0 r^3}\sum_{p=-2}^2 (-1)^{M_L} \sqrt{(2L+1)(2L'+1)}
\begin{pmatrix}
    L & 2 & L' \\
    0 & 0 & 0
\end{pmatrix}
\begin{pmatrix}
    L & 2 & L' \\
    -M_L & p & M_L'
\end{pmatrix}\nonumber \\
&\times \sum_{p_1,p_2}
\begin{pmatrix}
1 & 1 & 2 \\
p_1 & p_2 & p
\end{pmatrix}
\langle \eta_1 | T^1_{p_1}(\vec{d}_1) | \eta_1' \rangle \langle \eta_2 | T^1_{p_2}(\vec{d}_2) | \eta_2' \rangle
\label{eqn:ddimatrixelements}
\end{align}
\end{widetext}
where $\langle \eta_i | T^1_p(\vec{d}_{i}) | \eta_i'\rangle$ is a dipole matrix element for molecule $i$. Note that $H_\text{ddi}$ couples states with non-vanishing single-molecule dipole matrix elements, according to the partial wave selection rules $\Delta L = 0,\pm 2$; $\Delta M_L = 0,\pm 1, \pm 2$, $L= 0 \not\leftrightarrow L'=0$.

Matrix elements of the dipole operator can be found elsewhere, e.g. in Ref.~\cite{Hirota1985}. However, it will be useful in the discussion below to calculate matrix elements in a simple symmetric top basis, $|N \ell M_N\rangle$, which captures the essential physics of the dipolar interactions, with the electron and nuclear spins being spectator degrees of freedom. The dipole matrix elements in this simplified basis are~\cite{Hirota1985}
\begin{align}
\langle \eta &| T^1_p(\vec{d}) | \eta' \rangle \equiv \langle N \ell M_N | T^1_p(\vec{d}) | N' \ell' M_N' \rangle \nonumber \\
&= d (-1)^{N-M_N}
\begin{pmatrix}
N & 1 & N' \\
-M_N & p & M_N'
\end{pmatrix}
\nonumber \\
&\times (-1)^{N-\ell} \sqrt{(2N+1)(2N'+1)}
\begin{pmatrix}
N & 1 & N' \\
-\ell & 0 & \ell'
\end{pmatrix}
\label{eqn:dipolematrixelements}
\end{align}
where $|d|=1.465$~D is the molecule-frame electric dipole moment.

To calculate dipolar interaction potentials (e.g. Fig.~3 of the main text), we diagonalize the effective potential $V_{\rm eff} = H_\text{mol} + H_\text{ddi} + \frac{\hbar^2 L(L+1)}{2\mu r^2}$ over a range of intermolecular separations $r$, where the third term is the centrifugal barrier and $\mu$ is the reduced mass.

\begin{figure*}
\centering
\includegraphics[width=0.9\textwidth]{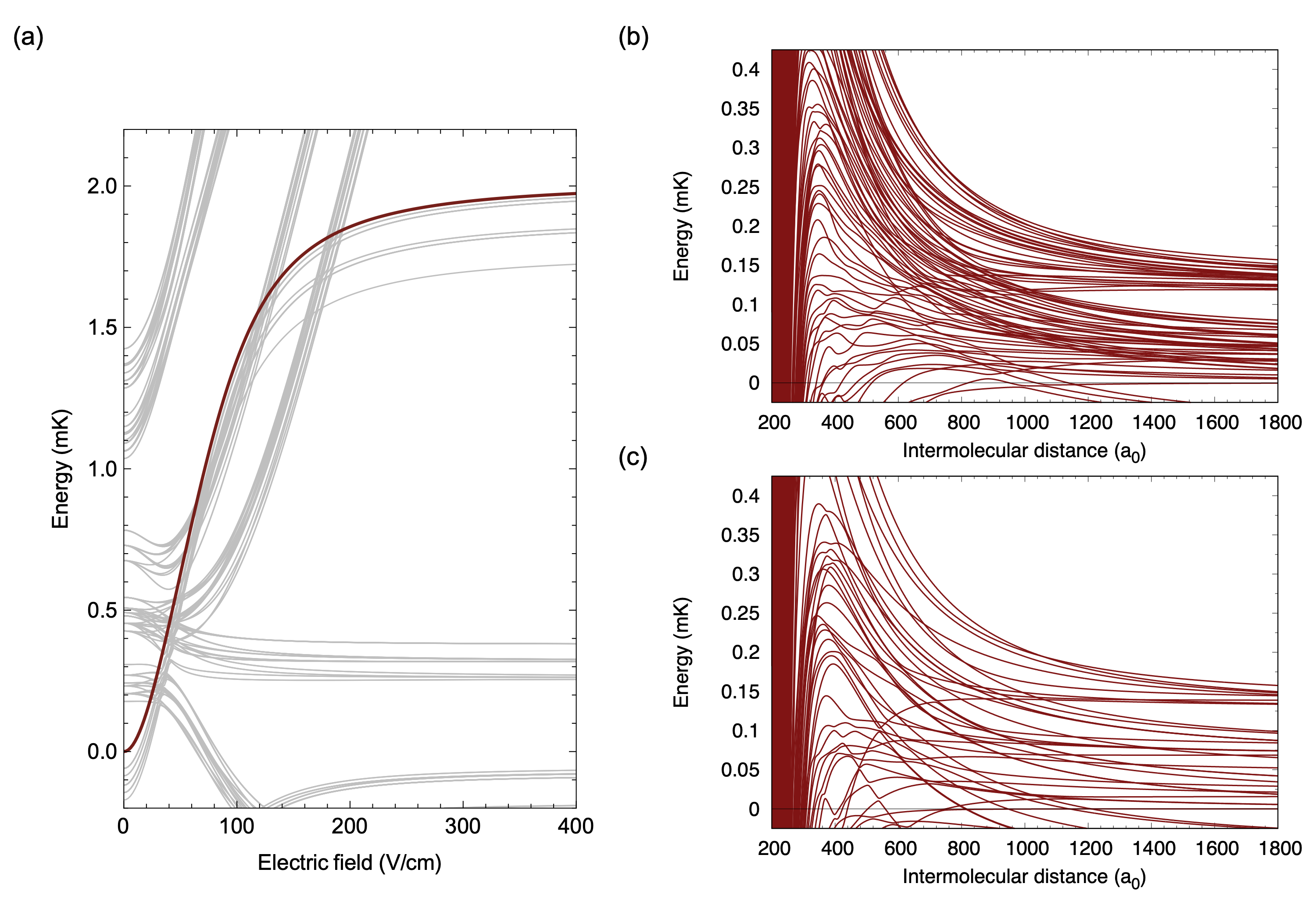}
\caption{Energies and adiabatic potential curves for molecules in the $d$ state at intermediate fields. (a) Asymptotic ($r\rightarrow \infty$) channel energy for the $|d\rangle|d\rangle$ channel (highlighted curve), showing that there are many channel crossings in the intermediate field range below $\mathcal{E}\approx 200$~V/cm. (b-c) Adiabatic curves of potentials for only $L=0, 2, 4$ partial waves at (b) $\mathcal{E} = 60$~V/cm and (c) $\mathcal{E} = 150$~V/cm. The energies of the collision partners include hyperfine structure and AC Stark shifts. For the potential curves, zero energy corresponds to the $J=1/2^-, F=1, M_F=\pm1$ states.}
\label{Adcurves}
\end{figure*}

Fig.~\ref{Adcurves} shows the non-interacting channel energy ($H_\text{mol}$ only) for two molecules in the $d$ state as a function of electric field, along with adiabatic dipolar interaction potentials at $\mathcal{E}=60$~V/cm and $\mathcal{E}=150$~V/cm, calculated as described in this section. The large density of states in this region may explain the resonant features seen in the data of Fig.~2 in the main text.

\subsection{Van der Waals coefficients}

For states with zero lab-frame dipole moment (e.g. at zero electric field), the leading order dipolar interaction 
is a 2nd-order, van der Waals (vdW) type interaction. In 2nd-order perturbation theory, the shift of channel $|\eta L M_L\rangle$ due to this mechanism is
\begin{align}
&U_\text{vdW} = \nonumber \\
&\sum_{\eta' L' M_L'} \frac{\langle \eta L M_L|H_\text{ddi}|\eta' L' M_L'\rangle \langle \eta' L' M_L' | H_\text{ddi} | \eta L M_L\rangle}{E_{\eta}^{(0)} - E_{\eta'}^{(0)}}
\label{eqn:vdWsum}
\end{align}
where the sum is over all channels to which $H_\text{ddi}$ couples the initial state, and $E^{(0)}_\eta$ is the asymptotic energy of channel $\eta$.

\subsubsection{Vibrational ground state}

The simplest case is to consider two molecules in the $N=0$ rotational level of the $\widetilde{X}(000)$ vibronic ground state, which does not have parity-doubling. For simplicity, we will ignore fine and hyperfine structure due to the electron and nuclear spin, which are spectator degrees of freedom. For the initial channel $|N_1=0, N_2=0, L=0, M_L = 0\rangle$, the leading interaction arises from 2nd order coupling to the $|N_1=1, M_{N_1}, N_2=1, M_{N_2}, L=2,M_L\rangle$ manifold, an energy $4B$ above. The single-molecule dipole matrix elements (eqn.~\ref{eqn:dipolematrixelements}) are $\langle N=0, M_N=0 | d | N=1, M_N \rangle = d/\sqrt{3}$, and summing the DDI matrix elements (eqn.~\ref{eqn:ddimatrixelements}) over all magnetic quantum numbers gives
\begin{widetext}
\begin{align*}
\sum_{M_{N_1}',M_{N_2}',M_L'} &|\langle N_1=0, M_{N_1}=0, N_2 = 0, M_{N_2} = 0, L=0, M_L = 0 | H_\text{ddi} | N_1'=1, M_{N_1}', N_2' = 1, M_{N_2}', L'=2, M_L' \rangle |^2 \\
&= \frac{2}{3}\left(\frac{d^2}{4\pi \epsilon_0 r^3} \right)^2
\end{align*}
\end{widetext}

Therefore, the van der Waals interaction between two molecules in $N=0$ is

\begin{align}
U_\text{vdW}(r) &= -\frac{2}{3}\left(\frac{d^2}{4\pi \epsilon_0 r^3} \right)^2 \frac{1}{4B} \nonumber \\
&= -\frac{1}{6B}\frac{d^4}{(4\pi \epsilon_0)^2} \frac{1}{r^6} \equiv -\frac{C_6}{r^6},
\end{align}
where $C_6 = (d^2/4\pi \epsilon_0)^2/(6B)$ is the vdW coefficient.

For the $|N_1=1,M_{N_1},N_2=1,M_{N_2}\rangle$ channels studied in the experiment, each individual channel experiences a slightly different potential. However, when averaged over all $M_{N_1}$ and $M_{N_2}$, $C_6$ is the same as in $|N_1=0, N_2=0\rangle$. For purposes of the universal loss calculations (below), we ignore the anisotropy and assume a single value $C_6 = (d^2/4\pi \epsilon_0)^2/(6B)$ for collisions of molecules in $\widetilde{X}(000)(N=1)$.

\subsubsection{Vibrational bending mode}

We now consider the $\widetilde{X}(010)$ vibrational bending mode. Our experimental measurements were in the lowest-lying rotational state, $N=1$. In this state, the 2nd-order perturbation sum, eqn.~\ref{eqn:vdWsum}, is dominated by coupling to nearby parity doublet states, and we can neglect coupling to higher rotational levels. Once again, we start by ignoring the electron and nuclear spin, and calculate matrix elements of $H_\text{ddi}$ using the parity basis states $|N^\pm,M\rangle \equiv 1/\sqrt{2}(|N,\ell,M\rangle \pm (-1)^{N-\ell}|N,-\ell,M\rangle$).
For the $|N_1=1^-, N_2 = 1^-\rangle$ channel, there is an attractive vdW interaction due to coupling to the $|N_1=1^+, N_2 = 1^+\rangle$ channel an energy $4q_\ell$ above.
Depending on the projection quantum numbers $M_{1}$ and $M_{2}$, there are four unique values for the vdW interaction strength from eqn.~\ref{eqn:vdWsum}:
\begin{widetext}
\begin{align*}
U_{00}(r) &= -\frac{1}{4q_\ell}\sum_{M_{1}',M_{2}',M_L'} |\langle M_{1}=0, M_{2} = 0, L=0, M_L = 0 | H_\text{ddi} | M_{1}', M_{2}', L'=2, M_L' \rangle |^2 = -\frac{7}{160 q_\ell}\left(\frac{d^2}{4\pi \epsilon_0 r^3} \right)^2 \\
U_{11}(r) &= -\frac{1}{4q_\ell}\sum_{M_{1}',M_{2}',M_L'} |\langle M_{1}=\pm1, M_{2} = \pm1, L=0, M_L = 0 | H_\text{ddi} | M_{1}', M_{2}', L'=2, M_L' \rangle |^2 = -\frac{1}{20q_\ell}\left(\frac{d^2}{4\pi \epsilon_0 r^3} \right)^2 \\
U_{10}(r) &= -\frac{1}{4q_\ell}\sum_{M_{1}',M_{2}',M_L'} |\langle M_{1}=\pm1, M_{2} = 0, L=0, M_L = 0 | H_\text{ddi} | M_{1}', M_{2}', L'=2, M_L' \rangle |^2 = -\frac{13}{320 q_\ell}\left(\frac{d^2}{4\pi \epsilon_0 r^3} \right)^2 \\
U_{-11}(r) &= -\frac{1}{4q_\ell}\sum_{M_{1}',M_{2}',M_L'} |\langle M_{1}=\pm1, M_{2} = \mp1, L=0, M_L = 0 | H_\text{ddi} | M_{1}', M_{2}', L'=2, M_L' \rangle |^2 = -\frac{11}{320q_\ell}\left(\frac{d^2}{4\pi \epsilon_0 r^3} \right)^2
\end{align*}
\end{widetext}
where we have omitted the labels $N_1 = N_2 = 1^-$ and $N_1' = N_2' = 1^+$ for brevity. Averaging over all channels gives $U_\text{vdW}^{N=1^-}(r) = (U_{00}+2U_{11}+4U_{10}+2U_{-11})/9 = -\frac{1}{24 q_\ell} \frac{d^4}{(4\pi \epsilon_0)^2}\frac{1}{r^6}$. Therefore,
\begin{equation}
    C_6 = \frac{d^2}{(4\pi \epsilon_0)^2}\frac{1}{24q_\ell}
\end{equation}
For $|N_1=1^+, N_2=1^+\rangle$, the vdW potential is repulsive with the same strength.

In CaOH, the spin-rotation interaction has a similar strength to the parity-doubling interaction, meaning that accurately calculating
$C_6$
requires accounting for the electron spin and including this interaction in the energy denominator. To do this, we evaluate matrix elements of
$H_\text{ddi}$ using single-molecule states $|N \ell S J M_J\rangle$. For brevity, below we will refer to parity-eigenstates in the $N=1$ manifold as $|J^\pm, M_J\rangle \equiv 1/\sqrt{2}(|N=1,\ell=1,S=1/2,J,M_J\rangle \pm (-1)^{N-\ell} |N=1,\ell=-1,S=1/2,J,M_J\rangle$.
We consider a few cases below.

In the lowest-energy, $|J_1=1/2^-, J_2=1/2^-\rangle$ channel at zero field, the vdW interaction is dominated by coupling to the $+$ parity channels $|1/2^+,1/2^+\rangle$, $|1/2^+,3/2^+\rangle$, and $|3/2^+,|3/2^+\rangle$, which are higher in energy by $4q_\ell$, $4q_\ell+3\gamma/4$, and $4q_\ell+3\gamma/2$, respectively, where $q_\ell$ is the $\ell$-doubling parameter and $\gamma$ is the spin-rotation parameter.
The contributions from each of these channels can be calculated by evaluating dipole matrix elements and averaging over all combinations of $M_{J_1}$ and $M_{J_2}$, as done above in the rigid rotor case. The result is:
\begin{align}
&U_\text{vdW}^{|1/2^-1/2^-\rangle}(r) \nonumber \\
&= -\left(\frac{2/27}{4q_\ell}+\frac{2\times 1/27}{4q_\ell+3\gamma/4}+\frac{1/54}{4q_\ell+3\gamma/2}\right)\left(\frac{d^2}{4\pi \epsilon_0 r^3} \right)^2 \nonumber \\
&\approx -0.85 \times \frac{1}{24 q_\ell} \left(\frac{d^2}{4\pi\epsilon_0}\right)^2 \frac{1}{r^6}
\end{align}

Average vdW interactions can be calculated in the same way for molecules in the other three $J$ manifolds in $N=1$, with the results being:
\begin{align}
&U_\text{vdW}^{|3/2^-3/2^-\rangle}(r) \nonumber \\
&= +\left(\frac{25/216}{4q_\ell}+\frac{2\times 5/216}{4q_\ell-3\gamma/4}+\frac{1/216}{4q_\ell-3\gamma/2}\right)\left(\frac{d^2}{4\pi \epsilon_0 r^3} \right)^2 \nonumber \\
&\approx -1.16 \times \frac{1}{24 q_\ell} \left(\frac{d^2}{4\pi\epsilon_0}\right)^2 \frac{1}{r^6}
\end{align}
\begin{align}
&U_\text{vdW}^{|1/2^+1/2^+\rangle}(r) \nonumber \\
&= +\left(\frac{2/27}{4q_\ell}+\frac{2\times 1/27}{4q_\ell-3\gamma/4}+\frac{1/54}{4q_\ell-3\gamma/2}\right)\left(\frac{d^2}{4\pi \epsilon_0 r^3} \right)^2 \nonumber \\
&\approx +1.36 \times \frac{1}{24 q_\ell} \left(\frac{d^2}{4\pi\epsilon_0}\right)^2 \frac{1}{r^6}
\end{align}
\begin{align}
&U_\text{vdW}^{|3/2^+3/2^+\rangle}(r) \nonumber \\
&= +\left(\frac{25/216}{4q_\ell}+\frac{2\times 5/216}{4q_\ell+3\gamma/4}+\frac{1/216}{4q_\ell+3\gamma/2}\right)\left(\frac{d^2}{4\pi \epsilon_0 r^3} \right)^2 \nonumber \\
&\approx +0.92 \times \frac{1}{24 q_\ell} \left(\frac{d^2}{4\pi\epsilon_0}\right)^2 \frac{1}{r^6}
\end{align}
The numerical prefactors can be interpreted as corrections to the rigid-body interaction strength due to the influence of the electron spin.

Note that since these expressions are derived in 2nd order perturbation theory, they deviate slightly from the potentials obtained by diagonalizing the full two-molecule Hamiltonian.

\section{Universal Loss Rates}
\label{sec:Universal}

\begin{figure*}
    \centering
    \includegraphics[width=1\textwidth]{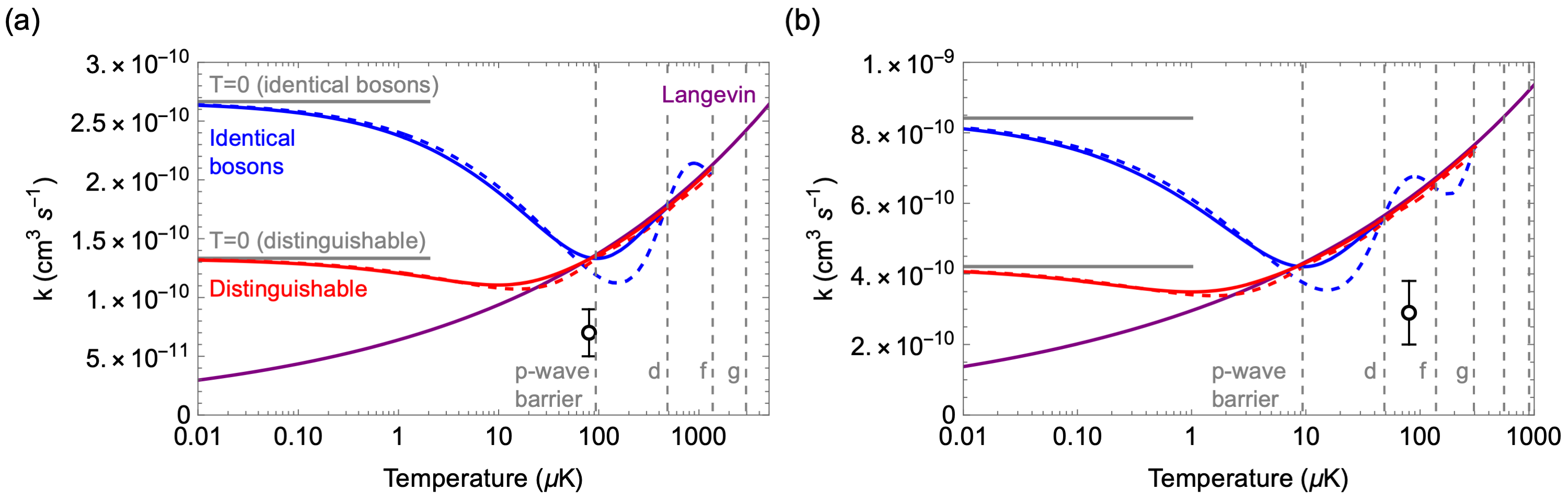}
    \caption{Calculated universal loss rate constants for (a) the $\widetilde{X}(000)$ state and (b) the $\widetilde{X}(010)(N=1^-)$ state. The low-temperature limit (horizontal grey lines) uses the expressions from Ref.~\cite{idziaszek2010universal}, and the high-temperature Langevin limit (purple curve) uses the expression from Ref.~\cite{jurgilas_2021}. Calculated rates in the intermediate temperature regime (blue and red curves) use the single-channel model from Ref.~\cite{frye2015cold}. Dashed curves are single-energy calculations, and solid curves are thermally averaged values. Data points are the experimental measurements from Fig.~1 in the main text.}
    \label{fig:UniversalLoss}
\end{figure*}

Universal loss rates for CaOH in the $\widetilde{X}(000)$ and $\widetilde{X}(010)(N=1^-)$ states are calculated assuming a long-range potential of the form $V(r) = -C_6/r^6$, where $C_{6,(000)} = \frac{1}{6B}\frac{d^4}{(4\pi \epsilon_0)^2}$ for $\widetilde{X}(000)$ and $C_{6,(010)} \approx 0.85 \times \frac{1}{24q_\ell}\frac{d^4}{(4\pi \epsilon_0)^2}$ for $\widetilde{X}(010)(N=1^-)$, as described above. The calculated universal loss rates are shown as a function of temperature in Fig.~\ref{fig:UniversalLoss}. At temperatures well below the p-wave barrier, the universal loss rate approaches a constant value given in Ref.~\cite{idziaszek2010universal}. At temperatures high enough for several partial waves to contribute, the universal loss is accurately described by a classical Langevin capture model~\cite{jurgilas_2021,bell2009ultracold}. At intermediate temperatures, universal loss rates may be calculated using a single-channel model based on analytic quantum defect theory (QDT), as described in Ref.~\cite{frye2015cold}.

For bosonic molecules in an incoherent mixture of $N$ hyperfine states, the rate equation describing collisional loss is
\begin{equation}
\dot{n}(t) = -k_\text{mixture} n(t)^2 = -\left(\frac{N+1}{N}k_e + \frac{N-1}{N} k_o\right) n(t)^2
\end{equation}
in terms of the rate constants summed over even ($k_e$) and odd ($k_o$) partial waves only~\cite{burau2024collisions}. Therefore, to a good approximation $k_\text{mixture} \approx  k_e + k_o$, i.e. we can use the calculated universal loss rate constant for distinguishable particles. Regardless, Fig.~\ref{fig:UniversalLoss} shows that at the 80~$\mu$K temperature in the experiment, the universal rate is similar for identical bosons and for distinguishable particles.

By comparing the 2nd-order estimates of $C_6$ with full diagonalization of the two-molecule Hamiltonian, we estimate that there is a 10-20\% uncertainty on the calculated universal rate constants using the values of $C_6$ described above. Additionally, the electronic contribution to the van der Waals interaction has not been included in these calculations. We can estimate this using the value for CaF, $C_6^\text{elec} \approx 2300 E_h a_0^6$~\cite{mukherjee2023shielding}, which is relatively small compared to the dipolar contributions discussed above, $C_{6,(000)} \approx 12000 E_h a_0^6$ and $C_{6,(010)} \approx 1.2 \times 10^6 E_h a_0^6$. Including this electronic contribution would lead to a $\sim10\%$ correction to the $\widetilde{X}(000)$ universal rate constant and a negligible change in the bending mode rate constant.

\section{Molecular scattering}

The collisional Hamiltonian of the two CaOH molecules is given by:
\begin{align}
\label{totHamiltonian}
H_{\rm tot} &= -\frac{\hbar^2}{2\mu}\frac{{\rm d}^2}{{\rm d}r^2} +  \frac{\hbar^2}{2\mu}\frac{{\bf L}^2}{r^2} + H_{\rm mol}   + H_{\rm ddi} + U_{\rm vdW} \nonumber \\
&\equiv -\frac{\hbar^2}{2\mu}\frac{{\rm d}^2}{{\rm d}r^2} + V_{\rm eff},
\end{align}
where ${\bf L}^2$ is the squared angular momentum of the rotation of the molecules about their center of mass; $\mu$ is the reduced mass of the pair of molecules; and $H_{\rm mol}$ is the Hamiltonian of the separated molecules, as described above.  All terms other than the radial kinetic energy are combined into an effective potential $V_{\rm eff}$.

To construct a basis set for scattering, the Hamiltonian $H_{X,i}$ (see eqn. \ref{eqn:heff}) of each molecule is diagonalized in the basis $|\eta_i \rangle$ described above, to produce a set of dressed eigenstates $|{\tilde \eta}_i \rangle$ for that molecule. A complete set of states is then given by the state of each molecule, times the partial wave angular momentum $|LM_L \rangle$.  Given that the molecules are identical bosons, we then construct channel functions symmetric under exchange of molecules, denoted schematically as
\begin{align}
|n \rangle = \frac{ |{\tilde \eta}_1 \rangle |{\tilde \eta}_2 \rangle + |{\tilde \eta}_2 \rangle |{\tilde \eta}_1 \rangle }{\sqrt{(1+\delta_{{\tilde \eta}_1{\tilde \eta}_2})}}|L M_L \rangle,
\end{align}
and with only even values of $L$ allowed.  In practice, we include even partial waves up to $L_{\rm max} = 16$.

The total wave function for two-body scattering is then
\begin{align}
| \Psi \rangle = \sum_{n=1}^{N_{\rm ch}} \psi_n(r) |n \rangle,
\end{align}
where $N_{\rm ch}$ is the number of scattering channels included in the calculation. In our case $N_{\rm ch} = 839$ for the $b$ and $f$ states, and $N_{\rm ch} = 1961$ for the $a,c,d$ and $e$ states.
Inserting this into the Schr\"odinger equation $(H_{\rm tot} - E_{\rm tot})|\Psi \rangle = 0$  and projecting onto alternative channel indices leads to the set of coupled radial Schr\"odinger equations
\begin{eqnarray}
\label{SchrEq}
\left( -\frac{\hbar^2}{2\mu}\frac{{\rm d}^2}{{\rm d}r^2} - E_{\rm tot} \right)\! \psi_n + \sum_{m=1}^{N_{\rm ch}} \langle n | V_{\rm eff} | m \rangle \,\psi_m = 0\,,
\end{eqnarray}
where $E_{\rm tot}$ is the total energy and $\psi_n$ is n-th component of the column vector $\psi(r)$.
Matrix elements $V_{\rm eff}$ are computed using the formulas given above, accounting for the transformation from the undressed $|\eta_i \rangle$ to the dressed $|{\tilde \eta}_i \rangle$ molecular bases.  The diagonalization of $V_{\rm eff}$ at each $r$ gives the adiabatic potential energy curves (see Fig. \ref{Adcurves}).

The coupled Schr\"odinger equations admit $N_{\rm ch}$ independent solutions, which, represented as column vectors, comprise a $N_{\rm ch} \times N_{\rm ch}$ matrix $M$.  
To solve eqn. \ref{SchrEq}, we propagate the logarithmic-derivative matrix $Y = M^{-1} dM/dr$ from an appropriate starting radius $r_0 =30\, a_0$
to an appropriate matching radius $r_m = 10\,000 \,a_0$. This propagation uses the logarithmic derivative matrix propagation method as developed by Johnson \cite{Johnson}.  This algorithm is efficient and stable, particularly in the presence of energetically closed channels, as we have here.

To account for short-range losses due to chemical reactions, we apply absorbing boundary conditions at $r=r_0$,
implying that any incident flux arriving at $r_0$ would be absorbed with unit probability.  To achieve this, the initial condition for the $Y$-matrix is diagonal, with diagonal elements equal to $Y_{nn} = -i k_n$, where $k_n = \sqrt{(2 \mu/\hbar^2)(E_{\rm tot}-V_{nn}(r_0)}$  is the local wave number in  channel $n$ at the initial radius $r_0$.

The matching radius $r_m$ is chosen to be so large that $V_{\rm eff}(r_m)$ is negligible.  At this point, the wave function is given by a linear combination of free-particle  solutions,
\begin{align}
    M_{nm} = h_{L_n}^- \delta_{nm} - h_{L_m}^+ S_{nm},
    \label{eq:S_def}
\end{align}
where $h^-$ and $h^+$  are incoming and outgoing spherical Hankel functions.  This expression defines the scattering matrix $S$.  Matching the logarithmic derivative of the right-hand side of (\ref{eq:S_def}) to the computed logarithmic derivative $Y$ determines the values of $S$.  Note that in general, $S$ is sub-unitary, accounting for the loss of incident flux to the short-range losses.

To compute scattering cross sections, we expand the channel notation slightly, writing each channel as a product of an internal states of the molecule pair, times the partial wave contribution, e.g., $|n \rangle = |i ,LM_L \rangle$.  Thereby cross sections for molecules initially in internal state $|i \rangle$ are computed as follows.  The collision energy $E_c = E_{\rm tot} - E_i$ is the energy above asymptotic threshold $E_i$.  This defines the incident wave number  $k_i =  \sqrt{2\mu E_{\rm c}}/\hbar$.  

\begin{figure*}
\centering
\includegraphics{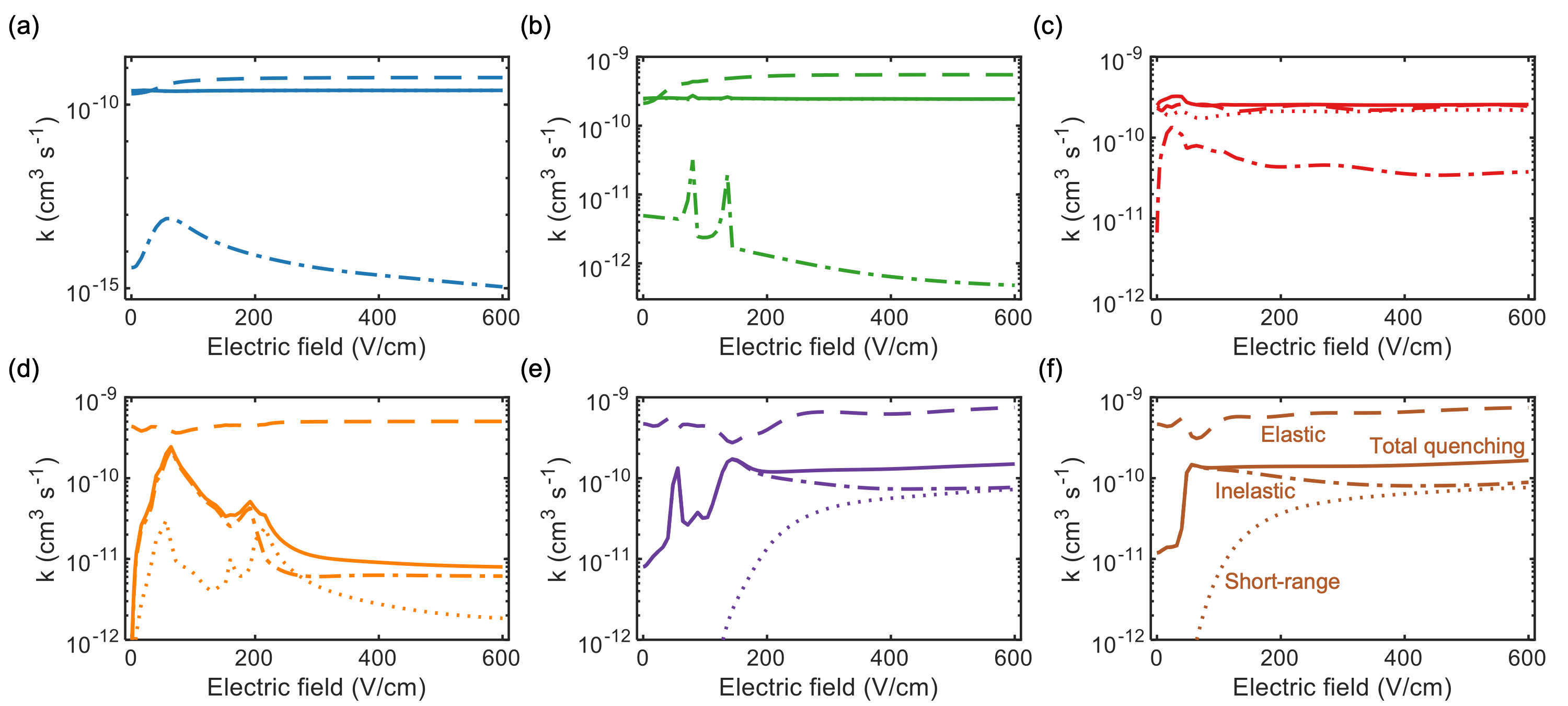}
\caption{Calculated collision rate coefficients for the $a$-$f$ manifolds vs. applied electric field, at a collision energy of $E_c = k_B \times 100$~$\mu$K. These are broken into  contributions from elastic collisions (dashed curves), inelastic collisions (dash-dotted curves), short-range absorption (dotted curves), and total collisional quenching (solid curves), which includes both inelastic collisions and short-range absorption.}
\label{fig:Calculated_contributions}
\end{figure*}

The cross sections for elastic, inelastic, and quenching collisions (the latter being the sum of inelastic, reactive, and absorption probabilities) are then given by \cite{Croft20}
\begin{align}
\label{Sif}
\sigma_{\rm el}(E_c) &= \frac{2\pi}{k_i^2}\sum_{LM_L} \sum_{L'M_L'} \big|1 - \langle i,LM_L |S| i,L'M_L'\rangle\big|^2 \\
\sigma_{\rm inel}(E_c) &= \frac{2\pi}{k_i^2}\sum_{f \ne i} \sum_{LM_L}  \sum_{L'M_L'} \big|\langle i,LM_L |S| f,L'M_L'\rangle\big|^2 \\
\sigma_{\rm qu}(E_c) &= \frac{2\pi}{k_i^2}\sum_{LM_L} \sum_{L'M_L'} \Big(\delta_{L,L'}\delta_{M_L,M_L'} - \nonumber \\
& \qquad \qquad \qquad \quad \big|\langle i,LM_L |S| i,L'M_L'\rangle\big|^2 \Big)
\end{align}
The scattering rate coefficient, for a single collision energy and one of the above processes, is defined as $K_{\rm proc}=v_i\sigma_{\rm proc}$, where $v_i = \sqrt{2E_c/\mu}$ is the velocity of the incident collision.

The calculated scattering rate coefficients are shown at a collision energy of $E_c = k_B \times 100$~$\mu$K in Fig.~\ref{fig:Calculated_contributions}. Individual contributions from elastic, inelastic, short-range absorption, and total quenching are shown. The elastic and quenching rate coefficients are also plotted in Fig.~2 of the main text.

%

\end{document}